\documentclass{aa}  

\usepackage{graphicx}
\usepackage{txfonts}
\usepackage{amsmath}
\usepackage{amssymb}
\usepackage{subcaption}
\usepackage{color} 
\usepackage{transparent}
\begin{document}
\title{Dissipative instabilities in a partially ionised prominence plasma slab: II. The effect of compressibility.}

\author{ J. F. Mather\inst{1}, I. Ballai\inst{1},   \and R. Erd\'elyi\inst{1,2},}
\institute{Solar Physics and Space Plasma Research Centre (SP$^2$RC), School of Mathematics and Statistics, The University of Shef{}field, Shef{}field,
UK, S3 7RH, email: {\tt jfmather1@sheffield.ac.uk}
\and Department of Astronomy, E\"otv\"os Lor\'and University, Pl\'azm\'any P. s\'et\'any 1/A Budapest, H-1117, Hungary}

\abstract{This present study deals with the dissipative instability that appears in a compressible partially ionised plasma slab embedded in a uniform magnetic field, modelling the state of the plasma in solar prominences. In the partially ionised plasma, the dominant dissipative effect is the Cowling resistivity. The regions outside the slab (modelling the solar corona) are fully ionised, and the dominant mechanism of dissipation is viscosity. Analytical solutions to the extended magnetohydrodynamic (MHD) equations are found inside and outside of the slab and solutions are matched at the boundaries of the slab. The dispersion relation is derived and solutions are found analytically in the slender slab limit, while the conditions necessary for the appearance of the instability is investigated numerically for the entire parameter space. Our study is focussed on the effect of the compressibility on the generation and evolution of instabilities. We find that compressibility reduces the threshold of the equilibrium flow, where waves can be unstable, to a level that is comparable to the internal cusp speed, which is of the same order of flow speeds that are currently observed in solar prominences. Our study addresses only  the slow waves, as these are the most likely perturbations to become unstable, however the time-scales of the instability are found to be rather large ranging from $10^5$-$10^7$ seconds. It is determined that the instability threshold is further influenced by the concentration of neutrals and the strength of the viscosity of the corona. Interestingly, these two latter aspects have opposite effects. Our numerical analysis shows that the interplay between the equilibrium flow, neutrals and dispersion can change considerably the nature of waves.
Despite employing a simple model, our study confirms the necessity of consideration of neutrals when discussing the stability of prominences under solar conditions.   }
\titlerunning{Dissipative instabilities in a partially ionised prominence plasma slab.}
\authorrunning{J. F. Mather. et al}
\maketitle
\section{Introduction}
Prominences are some of the most dynamical and, therefore, scientifically challenging structures that occur in the solar atmosphere. Suspended above the chromosphere and situated within the coronal plasma, they are made up of dynamical threads tracing the magnetic field lines that outline their structure. High-resolution H$\alpha$ observations within the past few decades have shown a more precise structure for these fibrils, with an average width of 200 km and heights ranging between 3500-28000 km (e.g. \citealp{2006ApJ...643L..65H} or \citealp{2011SSRv..158..237L} for more comprehensive reviews). Even though they are surrounded by coronal plasma, prominences have chromospheric temperatures, with lifetimes spanning from a few hours to a day in the case of active region prominences and three to 300 days in quiescent prominences (\citealp{2014LRSP...11....1P}). The fact that these structures do not collapse under gravity may seem to be surprising at first, but their stability is due to support from the magnetic tension forces, working against gravity.

Due to their chromospheric origin, prominences have a relatively low temperature compared to the surrounding coronal plasma, with temperatures ranging between 7500-9000 K (see e.g. \citealp{1998ASPC..150...23E}, \citealp{2005A&A...444..585L}, \citealp{2007Sci...318.1577O}). These low temperatures mean the plasma is likely only partially ionised, with the ratio of protons to neutral H-atoms in the range of 0.2-0.9 (\citealp{1990LNP...363.....R}). Therefore a one-fluid description may not be appropriate and multifluid models can describe the associated physics better. With the collisions between ions and neutrals, a resistivity acting only on perpendicular currents is introduced (\citealp{1965RvPP....1..205B}), namely the Cowling resistivity that is several orders of magnitude larger than the electron-generated Spitzer resistivity.

Instabilities within prominences are widely observed, the most obvious example being the turbulent eddies that are seen at the interface between the corona and prominence, which may be caused by instabilities associated with wave phenomena (\citealp{2010SoPh..267...75R}, \citealp{2010ApJ...716.1288B}). One explanation of these could be the Kelvin-Helmholtz instability (see, e.g. \citealp{1995SoPh..159..213N}, \citealp{2003SoPh..217..199T}, \citealp{2010A&A...516A..84Z}) that occurs when there is a shear flow between two different plasma layers (in relation to prominences see e.g. \citealp{2008ApJ...676L..89B}). A shear flow can also lead to resonant flow instabilities, which have been explained as a type of negative energy wave instability occurring at lower shear flow velocities than the Kelvin-Helmholtz instability (see, e.g. \citealp{1998A&A...332..786T}). The Rayleigh-Taylor instability, caused by heavier fluid sinking into lighter fluid under the force of gravity, is another instability that has been theorised to occur in prominences (see e.g. \citealp{2012ASPC..454..143R}). \cite{2012ApJ...746..120H} have suggested that the Rayleigh-Taylor instability generates some of the up-flows observed in prominences. 

Being such dynamic objects, plasma flows are ubiquitous in prominences. Bulk flows of around 10-70 km s${}^{-1}$ are observed within quiescent prominences (\citealp{1984A&A...136...81S}).  \cite{2008ApJ...676L..89B} observed, in both the Ca II H-line and H$\alpha$ band passes of Solar Orbiter Telescope (SOT), turbulent up-flows in quiescent prominences of an approximate constant speed of 20 km s${}^{-1}$. Active region prominences can exhibit even larger flows that can be up to a maximum of 200 km s${}^{-1}$. However, most of the observed flow speeds do not reach the threshold for the Kelvin-Helmholtz instability to occur. Therefore, other instabilities must be considered that require a lower flow threshold. 

The concept of negative energy waves (and the associated instabilities) has been studied previously in a solar physics context by many authors (e.g. \citealp{1996JPlPh..56..285R}, \citealp{1997SoPh..176..285J}, \citealp{1998A&A...332..786T}, \citealp{2003SoPh..217..199T}, \citealp{2015A&A...577A..82B}, amongst others). This instability, again, occurs in the  presence of shear flow, but some form of dissipation must be present in the system to account for this energy loss. Generally, the instability appears when the flow is great enough that the direction of propagation of the `backward propagating' wave changes. This tends to happen at flow velocities lower than those required for the Kelvin-Helmholtz instability (see, e.g. \citealp{1998JGR...10326573R}). 

\cite{2015A&A...577A..82B} considered the dissipative instability in the case of an interface between a viscous and incompressible coronal plasma and a resistive (Cowling resistivity), incompressible and  partially ionised prominence plasma. \cite{2017A&A...603A..78B} investigated a similar problem but now modelling the plasma as a slab, introducing dispersion. Both investigations found that, with the increase in the fraction of neutrals present, the value of the threshold velocity for the dissipative instability to occur increased. The present study is a natural extension to \cite{2017A&A...603A..78B} and generalisation to this problem by including compressibility with the aim to investigate how this affects the threshold of the dissipative instability. With this addition, we can study these instabilities in a more realistic solar context.

The paper is set out as follows: In Section 2, we present the equilibrium background and governing equations. Solutions governing the dynamics inside and outside the wave-guides are given, along with the necessary boundary conditions. In Section 3, the dispersion relation for the symmetric sausage waves is derived and the slender slab limit is taken for both the real and imaginary part of the frequency. In Section 4, analysis of the slender slab limit is completed and numerical solutions for the full dispersion relation are obtained. In Section 5, we summarise our results and conclude our research.

\section{Equilibrium}
\begin{figure}
			\includegraphics[height=7cm,width=0.5\textwidth]{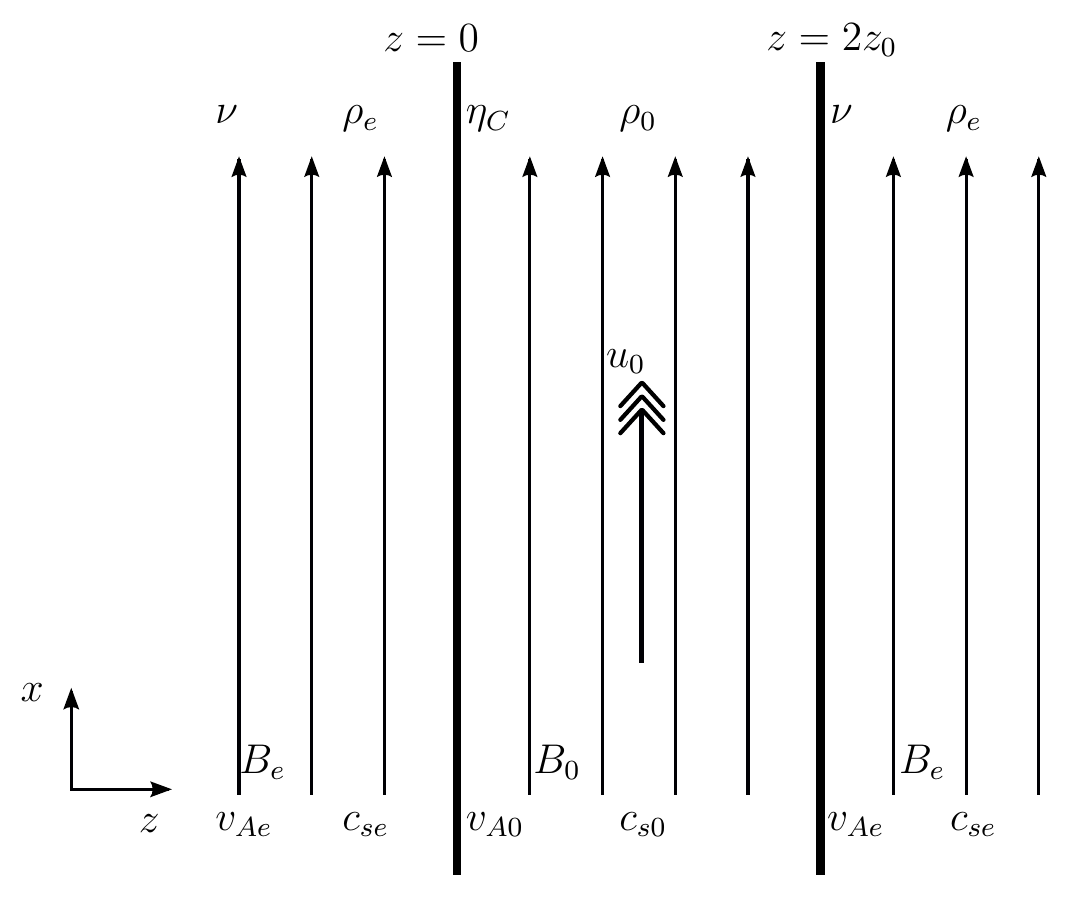}
			\label{fig1}
			\caption{Equilibrium configuration for the magnetic slab and its surrounding magnetic environment. The equilibrium flow inside the slab points in the positive $x$-direction.}
\end{figure}
We have considered a magnetised plasma slab of width $2z_0$ with boundaries at $z=0$ and $z=2z_0$ (see Figure 1), situated between to semi-infinite planes. The magnetic field in the three regions is uniform and parallel with the $x$-axis such that, ${\bf{B}}_0=B_0\hat{{\bf{x}}}$, where $\hat{{\bf{x}}}$ is the unit vector in the $x$-direction. A uniform background flow is present within the slab environment, parallel to the $x$-axis, that is ${\bf{v}}_0=u_0\hat{{\bf{x}}}$. The equilibrium temperatures, densities and pressures are assumed to vary on a scale that is much larger than the wavelength of the waves studied and are therefore considered constant functions in $z$, that is 

\begin{equation}
\begin{array}{lll}
T(z),& c_s(z),&\\
 v_A(z),&c_{T}(z),& u(z)
\end{array} =
\left\{
\begin{array}{llll}
T_e, & c_{se}& & z<0\;\;\mbox{\&}\;\; z>2z_0 ,
\\
 v_{Ae}, &  c_{Te},& 0,&\\
\\
T_0, & c_{s0},& & 0<z<2z_0 , \\
 v_{A0}, & c_{T0},& u_0,&
\end{array}
\right.
\label{Eq1}
\end{equation}  
where the indices $0$ and $e$ denote internal and external quantities of the slab, respectively, and $T_l$, $c_{sl}$, $v_{Al}$, $c_{Tl}$, $u_l$ ($l=0,e$) are the temperature, sound speed, Alfv\'en speed, tube speed (\mbox{$c_{Tl}=c_{sl}v_{Al}/(c_{sl}^2+v_{Al}^2)^{1/2}$}), and background  equilibrium bulk-flow speed inside the slab.

In our model, we have approximated the entire prominence by a slab and, therefore, the solar corona constitutes the external environment. Given the high probability that prominences are of chromospheric origin, we assumed that their temperature does not reach the threshold where hydrogen is fully ionised. In contrast, the surrounding corona, with its million degree temperature, is fully ionised. In partially ionised plasmas, the concept of transport mechanisms has to be treated differently, as the way momentum and energy is transported and dissipated is distinct from the formalism used in fully ionised plasmas. The reason for this is that the presence of neutrals and the collisions between heavy particles may change the nature and magnitude of these mechanisms (for a detailed discussion on the possible transport mechanisms in partially ionised plasmas see, e.g.  \citealp{2004A&A...422.1073K} or \citealp{2011A&A...529A..82Z}). Here, we have assumed that the ion-neutral collisional frequency is much larger than the harmonic motions of waves. Therefore the dynamics in the partially ionised plasma can be described within the framework of a single-fluid MHD. In addition, we assumed that there is strong thermal coupling between the ions, electrons a neutrals. Among all possible dissipative mechanisms, we concentrate on the processes that appear in the generalised Ohm's law, in particular, we deal with the classical and Cowling resistivity. The effects of these two resistivities are very different. The classical Spitzer resistivity is due to electrons and generates the dissipation of currents along magnetic field. Contrastingly, the Cowling resistivity is due to ions and creates a magnetic resistivity of currents that flow perpendicular to the ambient magnetic field. 

With these considerations, the dynamics in the partially ionised prominence are given by the system of linearised MHD equations
	\begin{equation}
	\frac{\partial p_1}{\partial t}
	+\left(
	{\bf{v}}_0\cdot\nabla\right)p_1 +
	\gamma p_0(\nabla\cdot{\bf{v_1}})=0,
	\label{energy1}
	\end{equation}
	\begin{align}\begin{split}
	\rho_0\left(\frac{\partial {\bf{v}}_1}{\partial t}
	+\left(
	{\bf{v}}_0\cdot\nabla\right){\bf{v}}_1\right)=&-\nabla p_1+
	(\nabla\times{\bf{B}}_1)\times\frac{{\bf{B}}_0}{\mu_0},
	\end{split}
	\label{momentum}
	\end{align}
	\begin{align}
	\begin{split} 
	\frac{\partial {\bf{B}}_1}{\partial t}+\left({\bf{v}}_0\cdot
	\nabla\right){\bf{B}}_1=&\left({\bf{B}}_0\cdot
	\nabla\right){\bf{v}}_1-
	{\bf{B}}_0\left(\nabla\cdot
	{\bf{v}}_1\right)+\eta\nabla^2{\bf{B}}_1\\
	&+\dfrac{\eta_C-\eta}{|{\bf{B}}_0|^2}\nabla\times\left\{\left[\left(\nabla\times{\bf{B}}_1\right)\times {\bf{B}}_0\right]\times{\bf{B}}_0\right\},
	\label{induction}
	\end{split}
	\end{align} 
\begin{align}
\nabla\cdot{\bf{B}}_1=0,
\label{solenoidal}
\end{align}
where all quantities with subscript $1$ denote small perturbations of background quantities. We note that the diamagnetic current that would be present in Eq. \eqref{induction} is neglected due to its weak dissipation and the fact we have considered a low plasma-beta. Here, $\rho_0=\rho_i+\rho_n$ so that the total background density, $\rho_0$, is always constant, with the ion density $\rho_i$ and the neutral density changing, depending on how ionised the plasma is. For a completely ionised plasma $\rho_i=\rho_0$ and for a fully neutral plasma $\rho_n=\rho_0$.

We assume all perturbations to be of the form
\[
f={\hat f}(z)\exp[i(kx-\omega t)],
\]
where $\hat{f}(z)$ is the amplitude of perturbations and the frequency, $\omega$, may be a complex quantity. Since the plasma is unbounded in the $x$ direction, we Fourier analyse the perturbations. In addition, instead of velocity components in the momentum equation \eqref{momentum}, we  use the Lagrangian displacement, $${\bf v}_1=\left(\dfrac{\partial}{\partial t}+
	{\bf{v}}_0\cdot\nabla\right){\boldsymbol \xi}.$$ 

Equations \eqref{energy1}-\eqref{solenoidal} along with Eq. \eqref{Eq1} can be then reduced to a  second order differential equation governing the Lagrangian displacement in the $z$-direction
\begin{equation}
	\xi_z''-\bar{M}_0^2\xi_z=0, \quad \bar{M}_0^2=-\dfrac{\left(\Omega^2-k^2c_{s0}^2\right)\left(\Omega^2-k^2\bar{v}_{A0}^2\right)}{(c_{s0}^2+\bar{v}_{A0}^2)(\Omega^2-k^2\bar{c}_{T0}^2)},
\label{governin}
\end{equation}
where the quantity $\bar{M}_0$ is the magnetoacoustic parameter, \mbox{$\bar{v}_{A0}^2=v_{A0}^2\left(1-i\eta_Ck^2/\Omega\right)$} is the modified Alfv\'en speed, while \mbox{$\bar{c}_{T0}^2=c_{s0}^2\bar{v}_{A0}^2/(c_{s0}^2+\bar{v}_{A0}^2)^2$ is} the associated cusp speed. Here, $v_{A0}=(B_0^2/\mu_0\rho_0)^{1/2}$ is the Alfv\'en speed and $c_{s0}=(\gamma p_0/\rho_0)^{1/2}$ is the sound speed of the internal medium, where $\gamma=5/3$ is the adiabatic index and $\mu_0$ is the magnetic permeability of free space. In the above equation, $\Omega=\omega-ku_0$ is the Doppler-shifted frequency.

The solutions inside the slab must be matched, at the boundaries of the slab with the solutions outside the slab. These boundaries must be stable to perturbations, that is any displaced fluid element on (or near) the boundary will not be advected by perturbations. This can be achieved only by assuming that the transversal component of the Lagrangian displacement is continuous and any other stresses that act in the perpendicular direction to the boundaries are matched by stresses on the other side of the boundary. That is why it is essential to use the Reynolds-Maxwell stress tensor to calculate the transversal stress component that is given by
	\begin{align}
		S_z=-\rho_0\dfrac{\left(
			\bar{v}_{A0}^2+c_{s0}^2\right)\left(
			\Omega^2-k^2\bar{c}_{T0}^2\right)}
		{\Omega^2-k^2c_{s0}^2}\xi_z'.
		\label{stressin}
	\end{align}

Let us now consider the dynamics in the surrounding coronal plasma. Due to the high temperatures of the corona ($>10^6K$), the plasma is fully ionised. Following on from the work by \cite{2017A&A...603A..78B}, we consider that the dominant dissipative mechanism that modifies the amplitude of waves is the viscosity. Given that the dynamics in the solar corona are driven mainly by magnetic forces, the viscosity is anisotropic and its value is given by the Braginskii's tensor present in the momentum equation. This tensor has five components, but by far the leading term (by about five orders of magnitude) is the first component (also called compressional viscosity) that is due to ions (see, e.g. \citealp{1965RvPP....1..205B}, \citealp{1996JPlPh..56..285R}). Therefore, the dynamics in the solar corona is described with the help of the linearised MHD equations
	\begin{equation}
	\frac{\partial p_1}{\partial t}+
	\gamma p_e(\nabla\cdot{\bf{v_1}})=0
	\label{energy2}
	\end{equation}
	\begin{align}\begin{split}
	\rho_e\frac{\partial {\bf{v}}_1}{\partial t}
	=&-\nabla p_1+
	(\nabla\times{\bf{B}}_1)\times\frac{{\bf{B}}_0}{\mu_0}\\
	&+\rho_e\nu\left\{{\bf{b}}({\bf{b}}\cdot\nabla)-\frac{1}{3}\nabla\right\}\left\{3{\bf{b}}\cdot\nabla
	({\bf{b}}\cdot{\bf{v}}_1)-\nabla\cdot{\bf{v}}_1\right\},
	\end{split}
	\label{momentum2}
	\end{align}
	\begin{align}
	\begin{split} 
	\frac{\partial {\bf{B}}_1}{\partial t}=\left({\bf{B}}_0\cdot
	\nabla\right){\bf{v}}_1-
	{\bf{B}}_0\left(\nabla\cdot
	{\bf{v}}_1\right),
	\qquad  \nabla\cdot{\bf{B}}_1=0,
	\label{induction2}
	\end{split}
	\end{align}  
where ${\bf b}={\bf B}_0/B_0$ is the unit vector in the direction of the equilibrium magnetic field, $\nu=\eta_0/\rho_e$ is the kinematic viscosity, and $\eta_0$ is the viscosity coefficient and its approximative value is given by
\[
\eta_0=\frac{\rho_e k_B T_e\tau_p}{m_p},
\]
where $T_e$ is the temperature of the external medium, $k_B$ is the Boltzmann constant, $\tau_p$ is the proton-proton collision time, and $m_p$ is the proton mass. For typical coronal values we can obtain that \mbox{$\eta_0\approx 5\times 10^{-2}$ kg m$^{-1}$ s$^{-1}$} (see, e.g. Hollweg 1985) 

In a similar way to the prominence plasma derivation, the MHD equations can be reduced to a single equation that describes the evolution of the transversal component of the Lagrangian displacement as 
\begin{equation}
	\xi_z''-M_e^2N_e^2\xi_z=0,\;\;\mbox{with}\;\; M_e^2=-\dfrac{\left(\omega^2-k^2c_{se}^2\right)\left(\omega^2-k^2v_{Ae}^2\right)}{(c_{se}^2+v_{Ae}^2)(\omega^2-k^2c_{Te}^2)},
\label{governout}
\end{equation}
and
\begin{equation}
	N_e^2=1+\dfrac{i\nu\omega}{3}
	\dfrac{\left(\omega^2-3k^2c_{se}^2\right)^2}{(c_{se}^2+v_{Ae}^2)\left(
\omega^2-k^2c_{Te}^2\right)\left(\omega^2-k^2c_{se}^2\right)}.
\end{equation}
Here, $v_{Ae}=(B_0^2/\mu_0\rho_e)^{1/2}$ is the Alfv\'en speed, $c_{se}=(\gamma p_e/\rho_e)^{1/2}$ is the sound speed of the external medium and ${c}_{Te}^2=c_{se}^2{v}_{Ae}^2/(c_{se}^2+{v}_{Ae}^2)^2$ is the external tube speed. In order to ensure that the interface between various regions is stable, the stress across the interfaces must be continuous, which now in the external region takes the form
	\begin{align}
	S_z=-\rho_e\dfrac{\left(
			v_{Ae}^2+c_{se}^2\right)\left(
			\omega^2-k^2c_{Te}^2\right)}
		{\omega^2-k^2c_{se}^2}\xi_z'
		+\rho_e\frac{i\omega\nu}{3}\dfrac{
			\left(
			\omega^2-3k^2c_{se}^2\right)^2}{\left(\omega^2-k^2c_{se}^2\right)^2}\xi_z',
\label{stressout}
	\end{align}
where $\xi_z$ is the solution of Eq. (\ref{governout}). Solving Eqs. \eqref{governin} and \eqref{governout} subject to the requirement that $\xi_z$ is finite as $|z|\to\infty$, gives the following general solutions within the slab and outside of the slab:
\begin{equation}
\xi_z =
\left\{
\begin{array}{ll}
C_1e^{-M_eN_e(z-2z_0)} & z>2z_0 ,
\\
&\\
C_2\sinh(\bar{M}_0z)+C_3\cosh(\bar{M}_0z) & 0<z<2z_0 , \\
&\\
C_4e^{M_eN_ez} & z<0 .
\end{array}
\right.
\label{solutions}
\end{equation} 
The constant coefficients $C_1\dots C_4$ can be determined after matching the solutions at the two boundaries. The general form of the solutions given by Eq. (\ref{solutions}), together with the governing equations inside and outside the slab will be used next to derive the dispersion relation for waves propagating inside the partially ionised magnetic slab.

\section{Dispersion relation}

The properties of waves propagating in the magnetic slab modelling the partially ionised prominence plasma can be studied with the help of the dispersion relation, that is the relation that gives the eigenvalues of the equations derived earlier in the presence of the boundary conditions that need to be imposed on the interfaces separating the two different media. According to the standard procedure (see e.g. \citealp{1996JPlPh..56..285R}), for a tangential discontinuity like in the present study, we require that the normal components of the displacement and the stress are continuous, that is their jump across the interface is zero. 

Depending on the choice of the solution inside the slab ($\sinh$ or $\cosh$) we find symmetric or antisymmetric modes, also labelled as sausage and kink modes, analogue to the fully ionised case, see e.g. \cite{1982SoPh...76..239E}. In what follows, we  concentrate on sausage modes, while the derivation of kink modes is similar. 

After equating the values of the displacements (imposing the kinematic boundary conditions) and stresses at $z=0$ and $z=2z_0$, we obtain the dispersion relation describing the propagation of sausage modes given by (considering large magnetic and viscous Reynolds numbers)
\begin{align}
	F_S(\omega,k)=F_{S0}(\omega,k)+iF_{S1}(\omega,k)=0, 
\label{sausagedisp}
\end{align}
where
\begin{align} 
	F_{S0}(\omega,k)=\rho_e\frac{D_{Ae}}{M_e}\tanh\left(M_0z_0\right)+\rho_0\frac{D_{A0}}{M_0},\label{Sausgeideal}
	\end{align}
is the real part of the dispersion relation, while 
\begin{align}
	\begin{split}
	F_{S1}=&
	\rho_eM_e
	\dfrac{\nu\omega}{6}\dfrac{D_{3se}^2}{D_{se}^2}\tanh\left(M_0z_0\right)\\
	&+\eta_C\frac{k^2v_{A0}^2}{2\Omega}\left[\rho_0M_0\dfrac{D_{m0}}{D_{s0}D_{A0}}\right. \\
&\quad\qquad\qquad\left.+\rho_e\dfrac{D_{Ae}}{M_e}\dfrac{\Omega^4}{D_{A0}D_{T0}}\left(1-\tanh^2\left(M_0z_0\right)\right)M_0z_0\right].
	\end{split}
\label{sausageimag}
	\end{align}
is the imaginary part of it. We note that Eq. \eqref{sausagedisp} was derived using a perturbation method considering terms multiplied by either $\nu^2$ or $\eta_C^2$ as negligibly small (as we assumed large Reynolds numbers). In the above equations the quantities have the following notation (for $l=0,e$)
\begin{align}
\begin{split} 
&D_{sl}=\Omega_l^2-c_{sl}^2k^2,\quad D_{Al}=\Omega_l^2-v_{Al}^2k^2,\\
&D_{Tl}=\left(c_{sl}^2+v_{Al}^2\right)\left(\Omega_l^2-c_{Tl}^2k^2\right),\\
& D_{3sl}=\Omega_l^2-3c_{sl}^2k^2,\quad
D_{ml}=\Omega_l^4-2k^2D_{Tl},\\
& \Omega_0=\omega-ku_0, \quad \Omega_e=\omega,\\
&M_0^2=-\dfrac{D_{s0}D_{A0}}{D_{T0}}.
\end{split}
\end{align}
 In general, the solutions of the dispersion relation \eqref{sausagedisp} are complex quantities (here, we assume that changes in the amplitude of waves are temporal effects, that is we assume that the frequency of waves is a complex quantity, while we set the wavenumber of waves as a real quantity). As the non-ideal ($F_{S1}$) part of Eq. \eqref{sausagedisp} is assumed to be small compared to the ideal part, that is $F_{S0}\gg F_{S1}$, the frequency of waves can be approximated as $\omega=\omega_0+i\omega'$, where $\omega_0$ is the solution of $F_{S0}(\omega_0,k)=0$ and $\omega'$ is the imaginary part of the frequency. By expanding Eq. \eqref{sausagedisp} around the solution of $F_{S0}(\omega_0,k)=0$ using a Taylor series, or applying the Cairns criterion (\citealp{1979JFM....92....1C}), we find that $\omega'$ may be approximated by 
 \begin{align}
		\omega'=-\dfrac{F_{S1}(\omega_0,k)}{\partial F_{S0}(\omega_0,k)/\partial\omega}.
\label{fullimag} 
\end{align}
Analytic progress in finding the solution to Eq. \eqref{sausagedisp} is still difficult as the equation is highly transcendental. In the long wavelength approximation (slender slab), both Eqs. \eqref{Sausgeideal} and \eqref{sausageimag} are considered in the limit $kz_0\to 0$, supposing that $M_iz_0\to 0$ (\citealp{1982SoPh...76..239E}). 

The method of dominant balance is then used to find a regular perturbation series representation of the solution. In this limit, the real part of the dispersion relation admits wave solutions that may, in general, be either surface ($M_i>0$)  or body ($M_i<0$) modes depending on values of background quantities (\citealp{1982SoPh...76..239E}). In our study, we use the term pseudo-body for one of the modes that  has a peculiar behaviour inside the slab, in the sense it does not have a spatially oscillatory structure in the $z$-direction but is a body mode as it satisfies the condition $M_i<0$ (for more information on these modes see, e.g. \citealp{2001A&A...377..330Z}, \citealp{doi:10.1063/1.1856931}, \citealp{2006SoPh..238...41E}, \citeyear{2007SoPh..246..101E}). In addition, there are also spatially oscillatory body-modes inside the slab which we term $n=1,2..$ body modes. These modes have nodes inside the slab. In the slender slab limit, up to first order in $(kz_0)^2$ and $kz_0$ for Eqs. \eqref{Fast} and \eqref{Slow}, respectively, the real solutions of the dispersion relation are
\begin{align}
		\omega_0\approx \pm kc_{se}\left\{1+
		\dfrac{\rho_e^2}{\rho_0^2}\dfrac{c_{se}^2\left(v_{Ae}^2-c_{se}^2\right)\left[\left(c_{se}-u_0\right)^2-c_{s0}^2\right]^2}{2\left(c_{s0}^2+v_{A0}^2\right)^2\left[\left(c_{se}-u_0\right)^2-c_{T0}^2\right]^2}\left(kz_0\right)^2\right\},
\label{Fast}	
\end{align} 
\begin{align}
		\omega_0\approx u_0k\pm kc_{T0}\left\{1+\frac{\rho_e}{\rho_0}\dfrac{\left[v_{Ae}^2-\left(\pm c_{T0}+u_0\right)^2\right]\left(c_{s0}^2-c_{T0}^2\right)}{\hat{M}_e\left(c_{s0}^2+v_{A0}^2\right)c_{T0}^2}kz_0\right\},
		\label{Slow}
	\end{align}
	where the upper signs describe the forward propagating waves, while the lower ones describe the backward propagating waves. We note that the first solution, given by Eq. \eqref{Fast}, only corresponds to a surface wave. Whereas the second solution, given by Eq. \eqref{Slow}, may be either slow or fast depending on the choice of the background. This solution is considered as the pseudo-body mode for the long wavelength approximation, as noted previously. In the above relation the parameter $\hat{M}_e$ is defined as
	\begin{align}
		\hat{M}_e^2=-\dfrac{\left[\left(\pm c_{T0}+u_0\right)^2-c_{se}^2\right]\left[\left(\pm c_{T0}+u_0\right)^2-v_{Ae}^2\right]}{(c_{se}^2+v_{Ae}^2)\left[\left(\pm c_{T0}+u_0\right)^2-c_{Te}^2\right]}.
	\end{align}
Obviously the quantity $\hat{M}_e$ has to be positive and this will impose a restriction on the domain of flows where our study remains valid. A simple analysis would reveal that waves will be able to propagate in the slab provided $|u_0\pm c_{T0}|<c_{Te}$ or 
\[
\mbox{min}(c_{Se},v_{Ae})<|u_0\pm c_{T0}|<\mbox{max}(c_{Se},v_{Ae})
.\]
Both Eqs. \eqref{Fast} and \eqref{Slow} show that waves are dispersive, that is their frequency depends on the wavelength of waves in a nonlinear way. Based on the leading order terms of the two equations, we can consider that the first equation (Eq. 20) describes the evolution of fast waves, while slow magnetoacoustic modes are described by the second expression (Eq. 21). For the case considered in the current study, we are going to investigate the mode that has the phase-speed approximately $c_{T0}$ (when $u_0=0$). We choose this mode over the mode given by Eq. \eqref{Fast} as the phase speed of this mode is affected linearly by the background flow (in the zeroth order approximation) and can thus change its direction of propagation, leading to the dissipative instability mentioned previously. The imaginary part of the frequency that is associated with the wave described by Eq.~(\ref{Slow})  is given by
	\begin{align}
\begin{split}
		\omega'=\mp &\left[\rho_r\dfrac{c_{s0}^4k^2\nu\left(\pm c_{T0}+u_0\right)\left[\left(\pm c_{T0}+u_0\right)^2-3c_{se}^2\right]^2}{6\left[\left(\pm c_{T0}+u_0\right)^2-c_{se}^2\right]^2c_{T0}\left(v_{A0}^2+c_{s0}^2\right)^2}\hat{M}_ekz_0\right. \\
& \left.\pm\dfrac{\eta_{C}v_{A0}^2c_{s0}^4k^2}{2c_{T0}^2\left(v_{A0}^2+c_{s0}^2\right)^2}\right],
\end{split}
		\label{imaginarypart}
	\end{align}
where $\rho_r=\rho_e/\rho_0$ is the relative density of our equilibrium and the Cowling resistivity, $\eta_C$ given by
\begin{equation}
\eta_C=\dfrac{v_{A0}^2m_n(2\mu-1)}{2\rho_0(1-\mu)\Sigma_{in}}\sqrt{\dfrac{\pi m_p}{k_BT_0}},
\end{equation}
where $m_n$ is the mass of a neutral atom, \mbox{$\Sigma_{in}=5\times10^{-15}$ cm${}^{2}$} is the ion-neutral collisional cross section and $\mu$ is the ionisation degree of the plasma given by $$\mu=\dfrac{1}{2-\xi_n},$$ where $\xi_n=\rho_n/\rho_0$ is the neutral fraction of the internal plasma. Using the range of ion to neutral H atom number density ratio ($\rho_i/\rho_n$), 0.2-0.9 given by \citet{1990LNP...363.....R}, we find a typical range for the ionisation degree $\mu$ from 0.65-0.85. In our calculations, we assume that \mbox{$\rho_0=5\times10^{-11}$ kg m${}^{-3}$}. Using coronal parameters we can estimate the magnetic Reynolds number, $R_m$ for this particular setup. Let us define
\begin{align}
\hat{\eta}_C=\dfrac{v_{A0}^2m_n}{2\rho_0\Sigma_{in}}\sqrt{\dfrac{\pi m_p}{k_BT_0}}, \quad \eta_C=\hat \eta_C\dfrac{2\mu -1}{1-\mu}.
\label{etahat}
\end{align}
We use a length scale of approximately $10^7$ m, which is a typical wavelength observed in prominences which tend to be in the range $10^6$-$10^8$ m. Several observational evidences support our choice for typical wavelength, for example \citet{1981SoPh...70..115M} reports on observed waves with wavelengths of $3.7\times 10^7$ m, \citet{1991A&A...243..501T}, \citet{1997SoPh..172..181M} and \citet{2002A&A...393..637T} observed waves in solar prominences with wavelengths $5.0\times 10^7$ m, $2.0\times 10^7$ m and $7.0\times 10^7$ m, respectively (for more examples see \citealp{2012LRSP....9....2A} and references therein). With typical prominence background wave velocities (i.e., the internal Alfv\'en and sound speeds) of $10^4$ m s${}^{-1}$, the magnetic Reynolds number is then given as:
\begin{align}
R_m\approx 10^4\dfrac{1-\mu}{2\mu-1}.
\end{align}
 The sign of the imaginary part of the frequency will determine whether a wave will be damped or amplified due to instabilities. According to the ansatz used in the present paper, $\omega'>0$ s${}^{-1}$ would mean that the wave is amplified, while waves will be damped in the opposite case when $\omega'<0$ s${}^{-1}$. 

\section{Dissipative instability}

In what follows, we are going to concentrate on the instabilities that arise due to the coupling between backward propagating waves inside the slab with the counter flow. These modes (for a particular combination of parameters) will have a positive imaginary part of the frequency, leading to an instability contrary to our natural physical intuition. The imaginary part of the frequency describing unstable behaviour is also connected to the dissipative processes.

\subsection{The slender-slab limit ($kz_0\ll1$)}\label{slenderslab}
 \begin{figure*}
	\begin{subfigure}[b]{0.5\textwidth}
		\includegraphics[height=7cm,width=\textwidth]{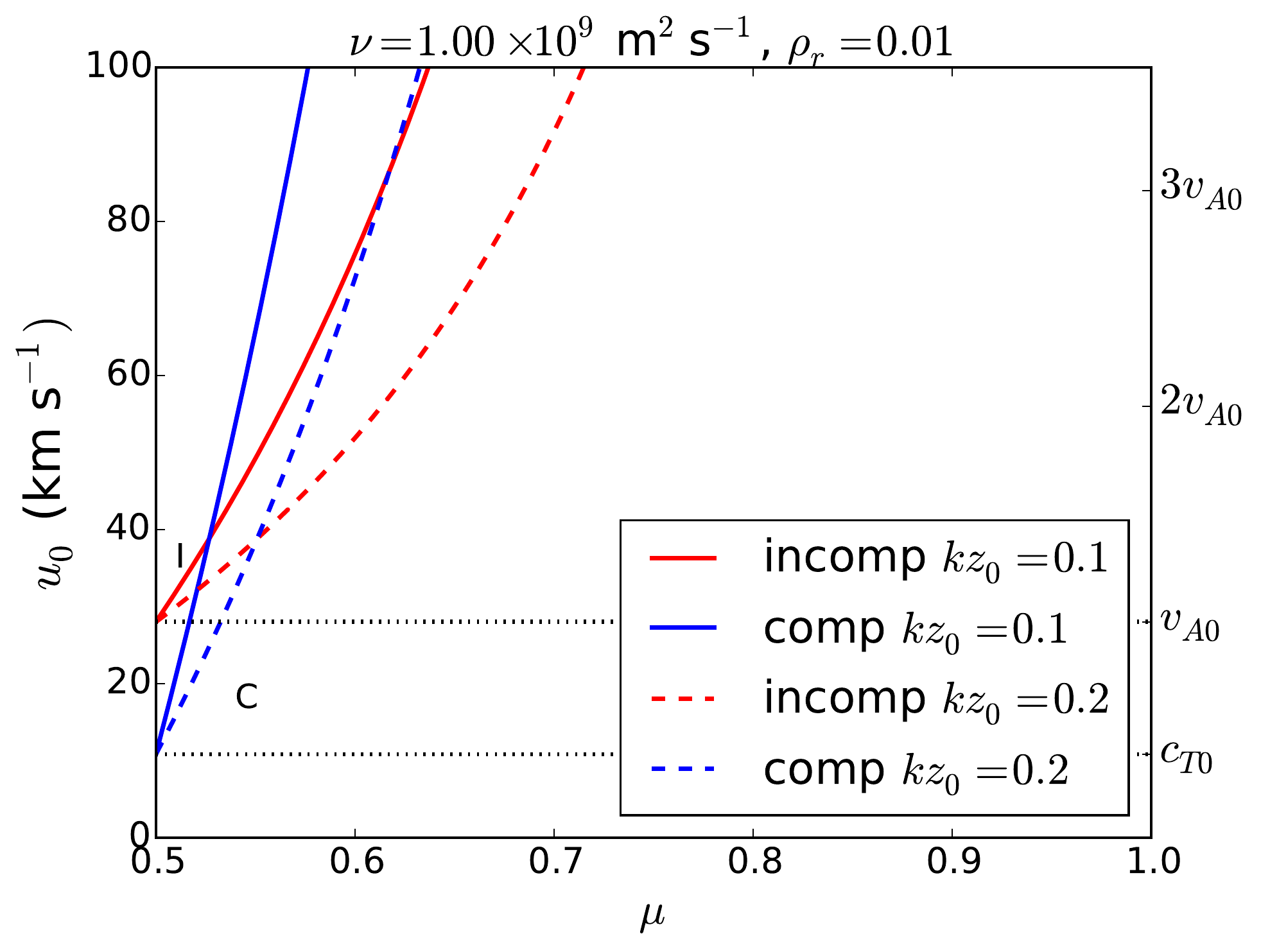}
		\caption{}
		\label{compression1}
			\end{subfigure}
	\begin{subfigure}[b]{0.5\textwidth}
		\includegraphics[height=7cm,width=\textwidth]{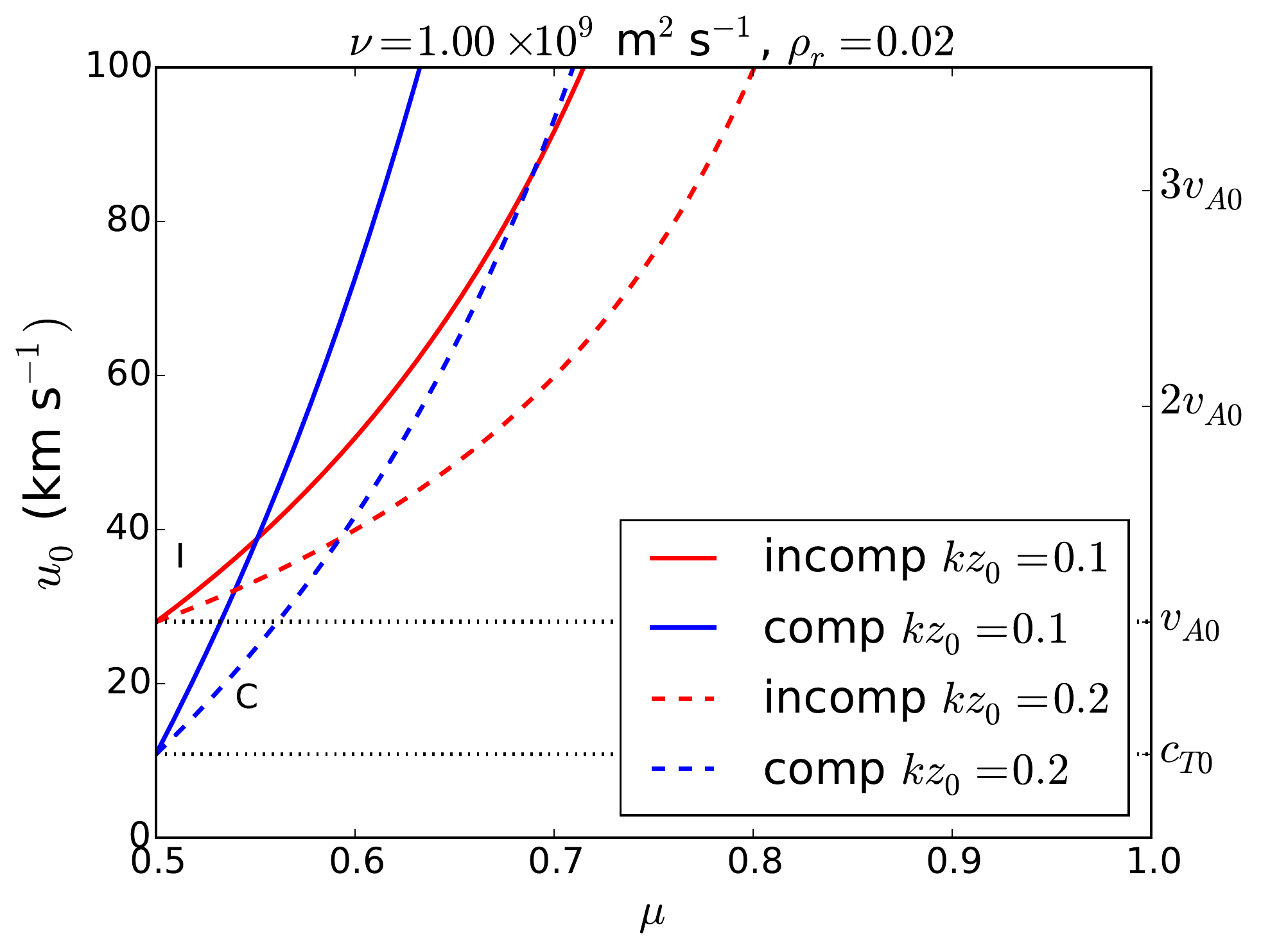} 
		\caption{}
		\label{compression2}
	\end{subfigure}
	\begin{subfigure}[b]{0.5\textwidth}
		\includegraphics[height=7cm,width=\textwidth]{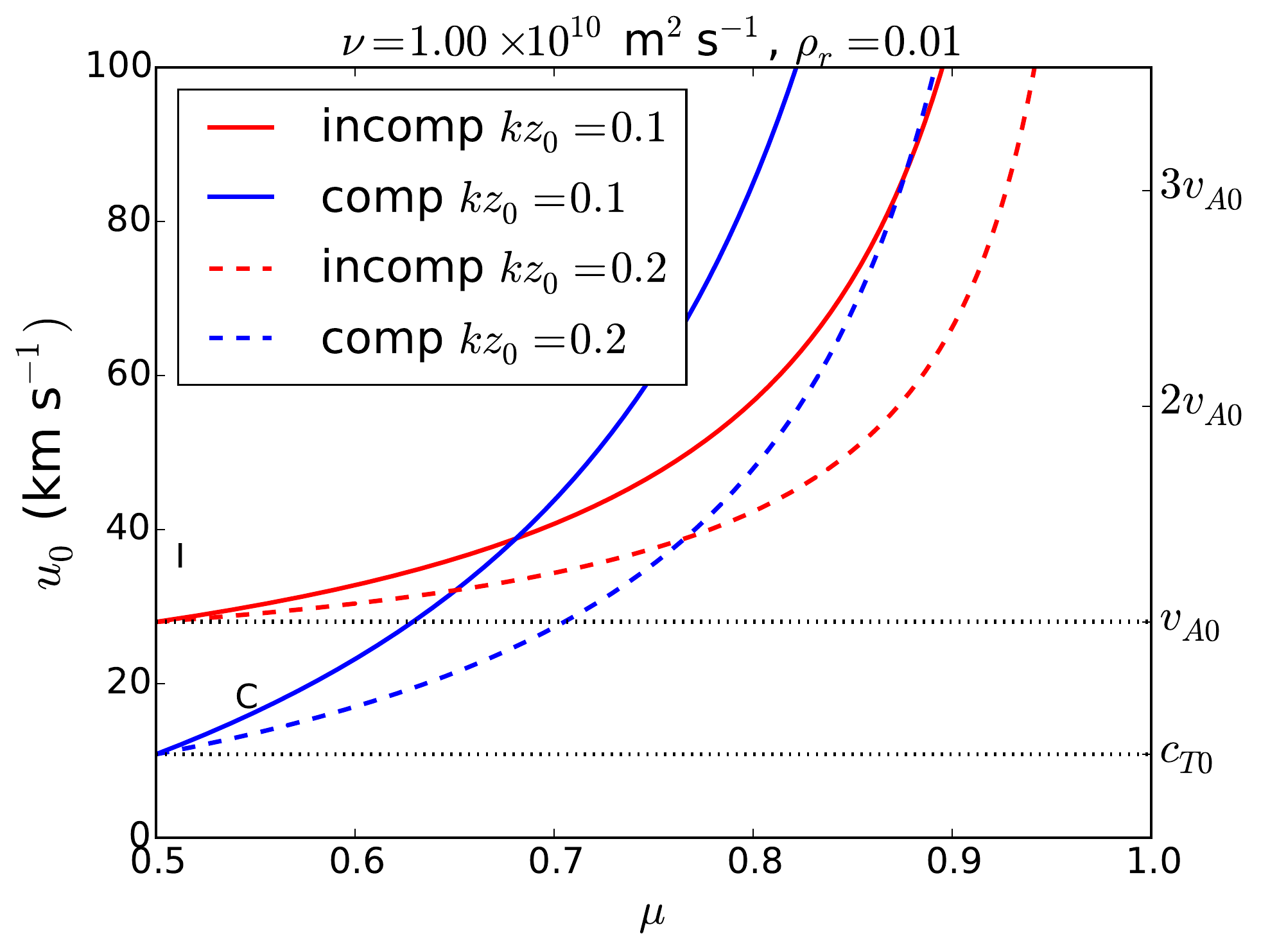}
		\caption{}
		\label{compression3}
			\end{subfigure}
	\begin{subfigure}[b]{0.5\textwidth}
		\includegraphics[height=7cm,width=\textwidth]{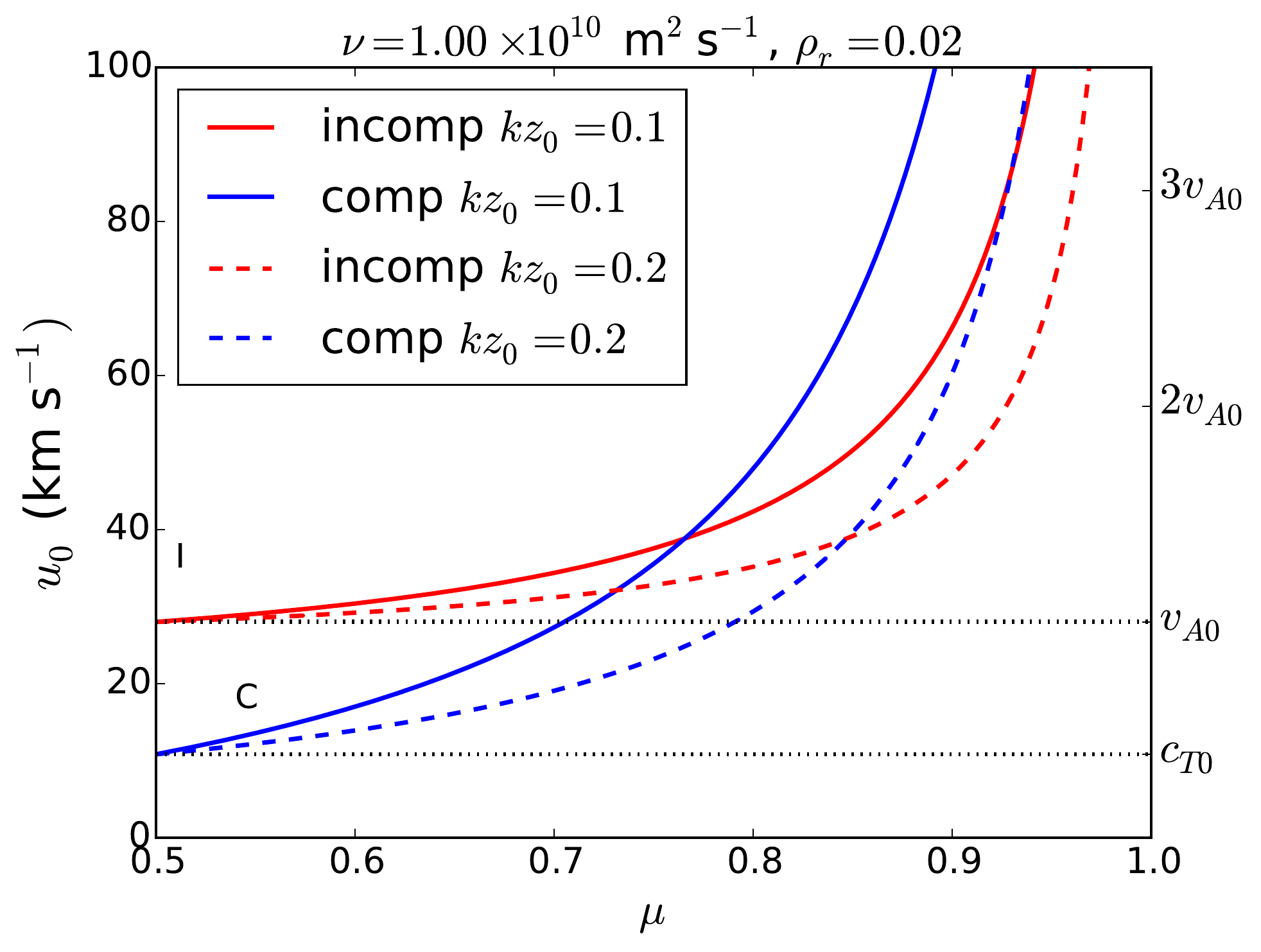} 
		\caption{}
		\label{compression4}
	\end{subfigure}
	\caption{Variation of the critical flow speed with respect to ionisation degree, $\mu$, for the compressible case (blue lines), given by Eq. \eqref{critflow}, and the incompressible case (red lines), given by Eq. \eqref{critflowincomp}: (a) and (b) both with $\nu=10^{9}$ m${}^2$s${}^{-1}$ but with $\rho_r=0.01$ and $\rho_r=0.02$ respectively and (c) and (d) both with $\nu=10^{10}$ m${}^2$s${}^{-1}$ but with $\rho_r=0.01$ and $\rho_r=0.02$ respectively. In each the solid lines (-----) indicate $kz_0=0.1$ and the dashed lines (- - -) indicate $kz_0=0.2$. The regions under the curves correspond to the stable regime.}
	\label{compression}
\end{figure*}
\begin{figure*}
	\begin{subfigure}[b]{0.5\textwidth}
		\includegraphics[height=7cm,width=\textwidth]{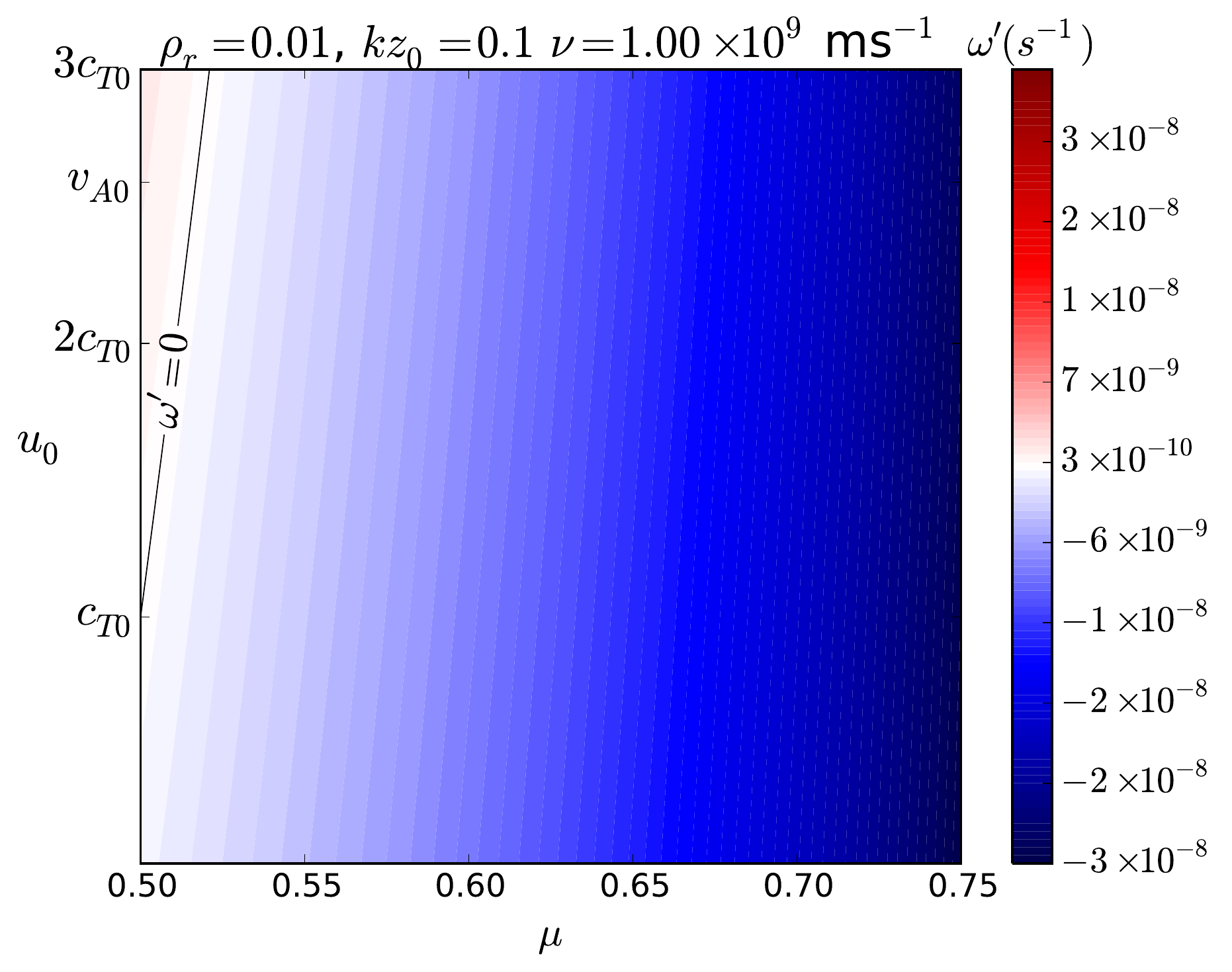}
		\caption{}
		\label{imagpart1}
			\end{subfigure}
	\begin{subfigure}[b]{0.5\textwidth}
		\includegraphics[height=7cm,width=\textwidth]{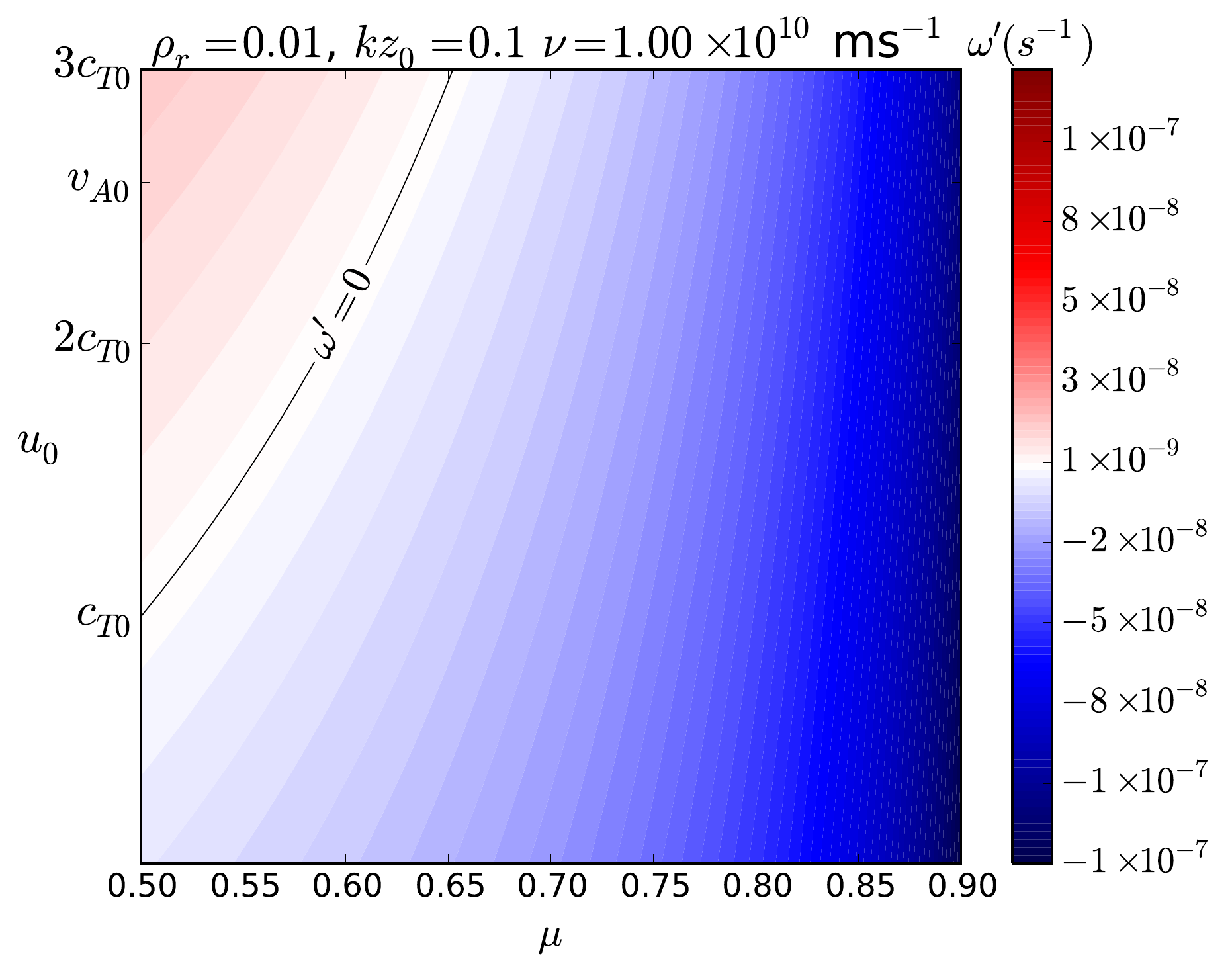} 	\caption{}
		\label{imagpart2}
	\end{subfigure}
\begin{subfigure}[b]{0.5\textwidth}
		\includegraphics[height=7cm,width=\textwidth]{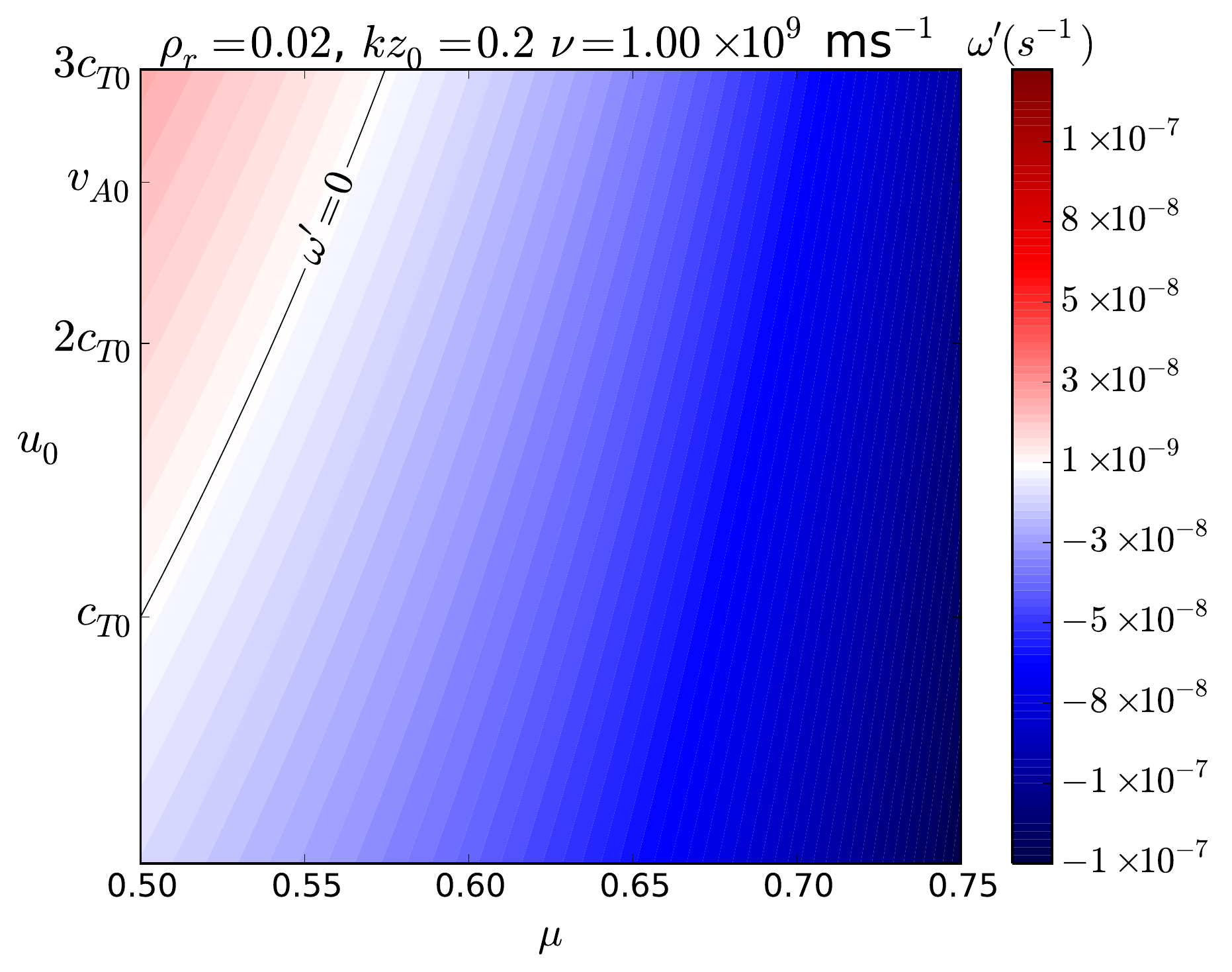}
		\caption{}
		\label{imagpart3}
			\end{subfigure}
	\begin{subfigure}[b]{0.5\textwidth}
		\includegraphics[height=7cm,width=\textwidth]{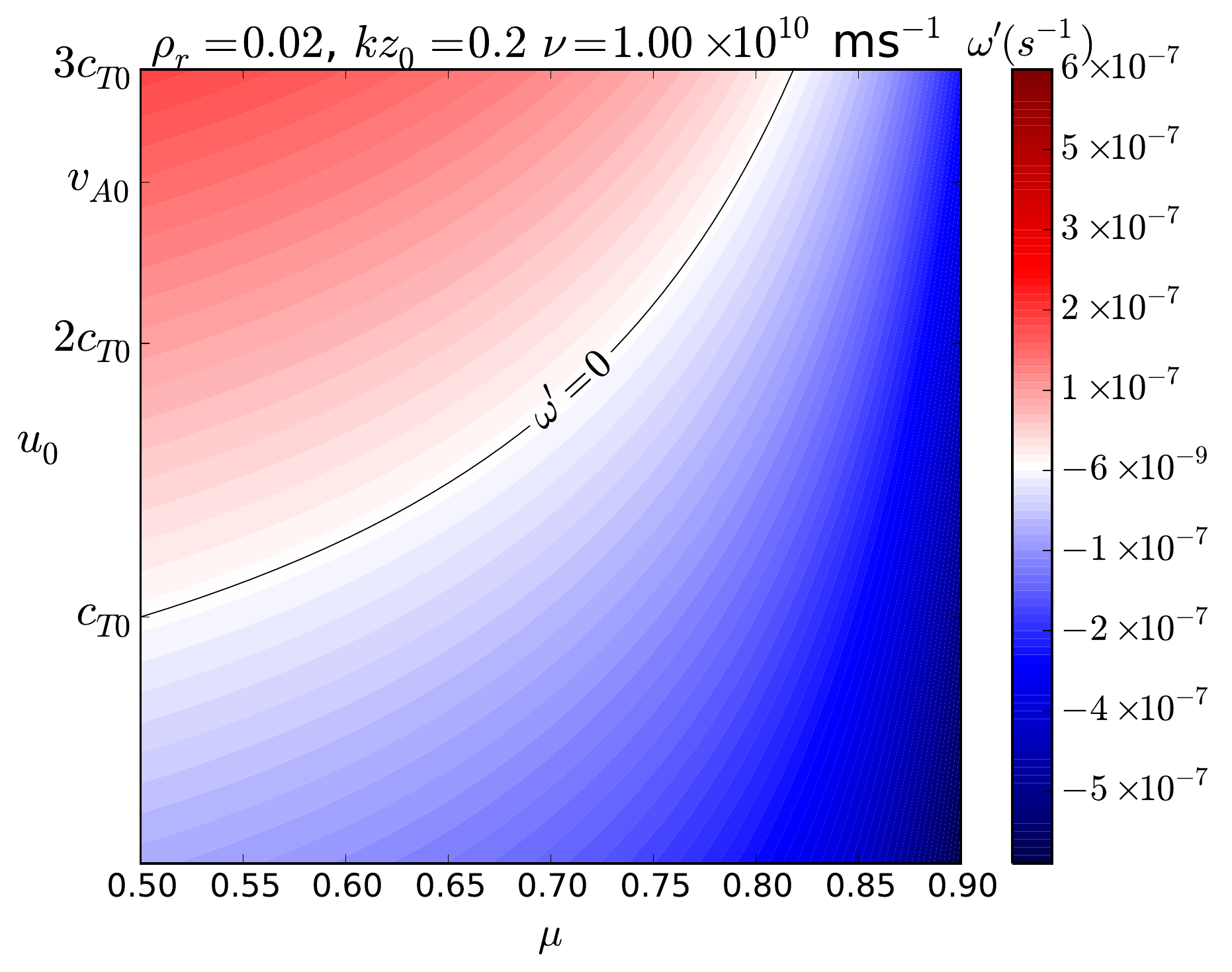} 	\caption{}
		\label{imagpart4}
	\end{subfigure}
	\caption{Contour plots showing the variation of $\omega'$ with respect to the internal background flow speed, $u_i$, and the ionisation degree $\mu$. Red indicates a positive value of $\omega'$ (amplification) and blue indicates a negative value of $\omega'$ (damping). The colour bar shows the numerical value of $\omega'$ and the contour labelled $\omega'=0$ indicates the transition between amplification and damping of the wave. Panels (a) and (b) both use the parameters \mbox{$\rho_r=0.01$}, \mbox{$kz_0=0.1$} and \mbox{$k=2\times10^{-7}$ m${}^{-1}$} but with  \mbox{$\nu=10^{9}$ m${}^2$s${}^{-1}$} and \mbox{$\nu=10^{10}$ m${}^2$s${}^{-1}$} respectively. In the panels (c) and (d) we used \mbox{$\rho_r=0.02$}, \mbox{$kz_0=0.2$} and  \mbox{$k=4\times10^{-7}$ m${}^{-1}$} along with  \mbox{$\nu=10^{9}$ m${}^2$s${}^{-1}$} and \mbox{$\nu=10^{10}$ m${}^2$s${}^{-1}$} respectively.  }
	\label{imagpart}
\end{figure*}
\begin{figure*}
	\begin{subfigure}[b]{0.5\textwidth}
		\includegraphics[height=7cm,width=\textwidth]{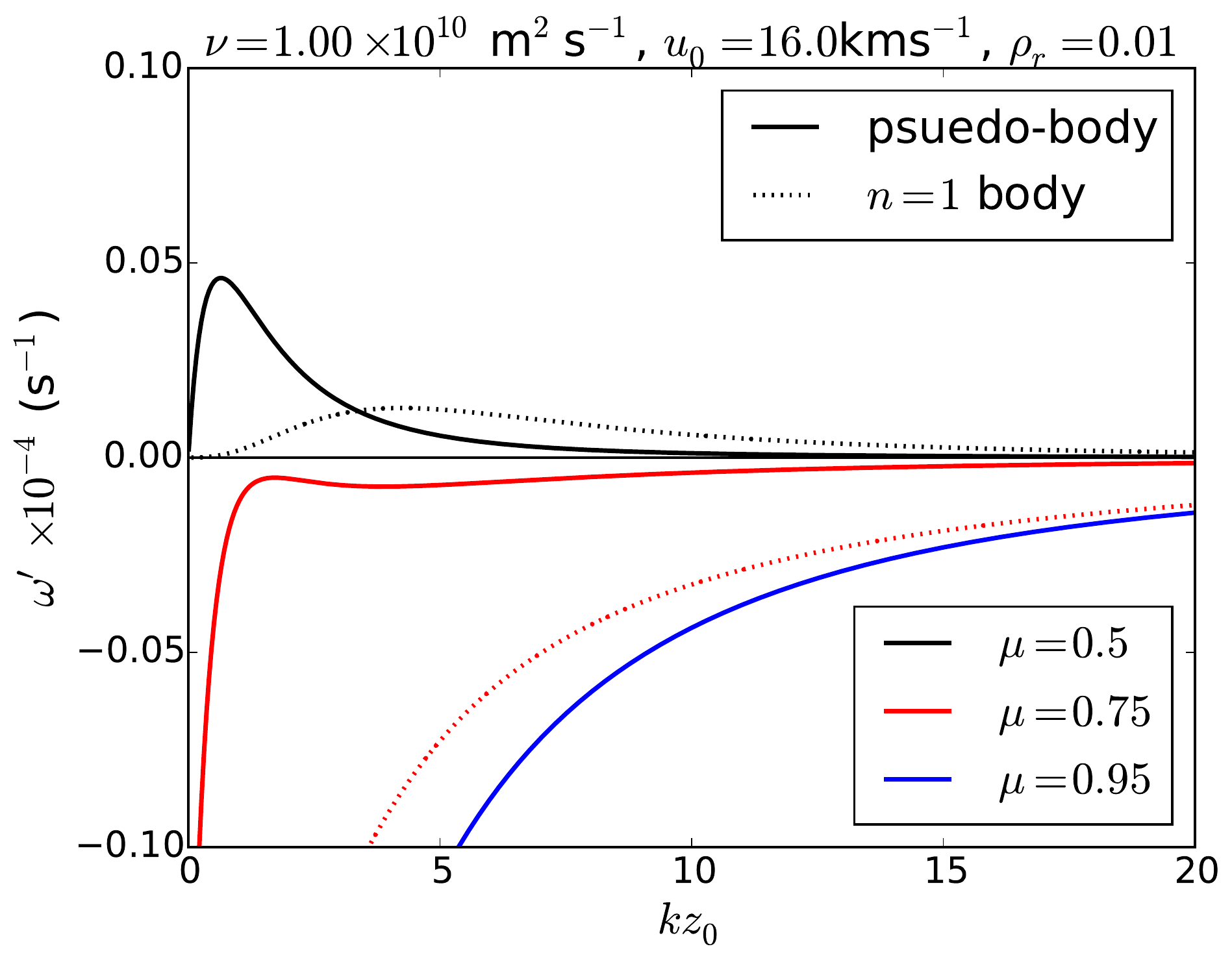}
		\caption{}
		\label{changek0011}
			\end{subfigure}
	\begin{subfigure}[b]{0.5\textwidth}
		\includegraphics[height=7cm,width=\textwidth]{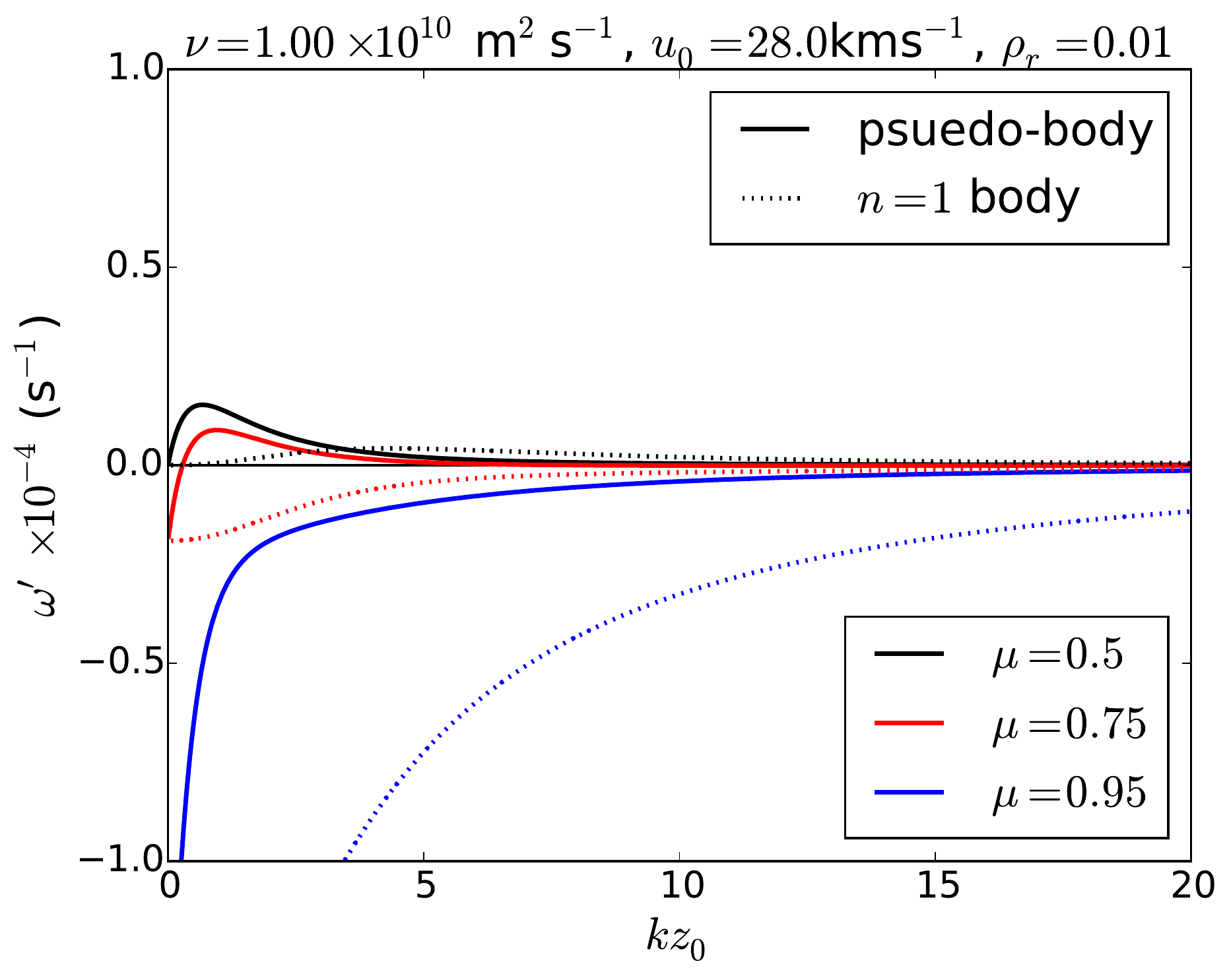} 	\caption{}
		\label{changek0012}

	\end{subfigure}
\begin{subfigure}[b]{0.5\textwidth}
		\includegraphics[height=7cm,width=\textwidth]{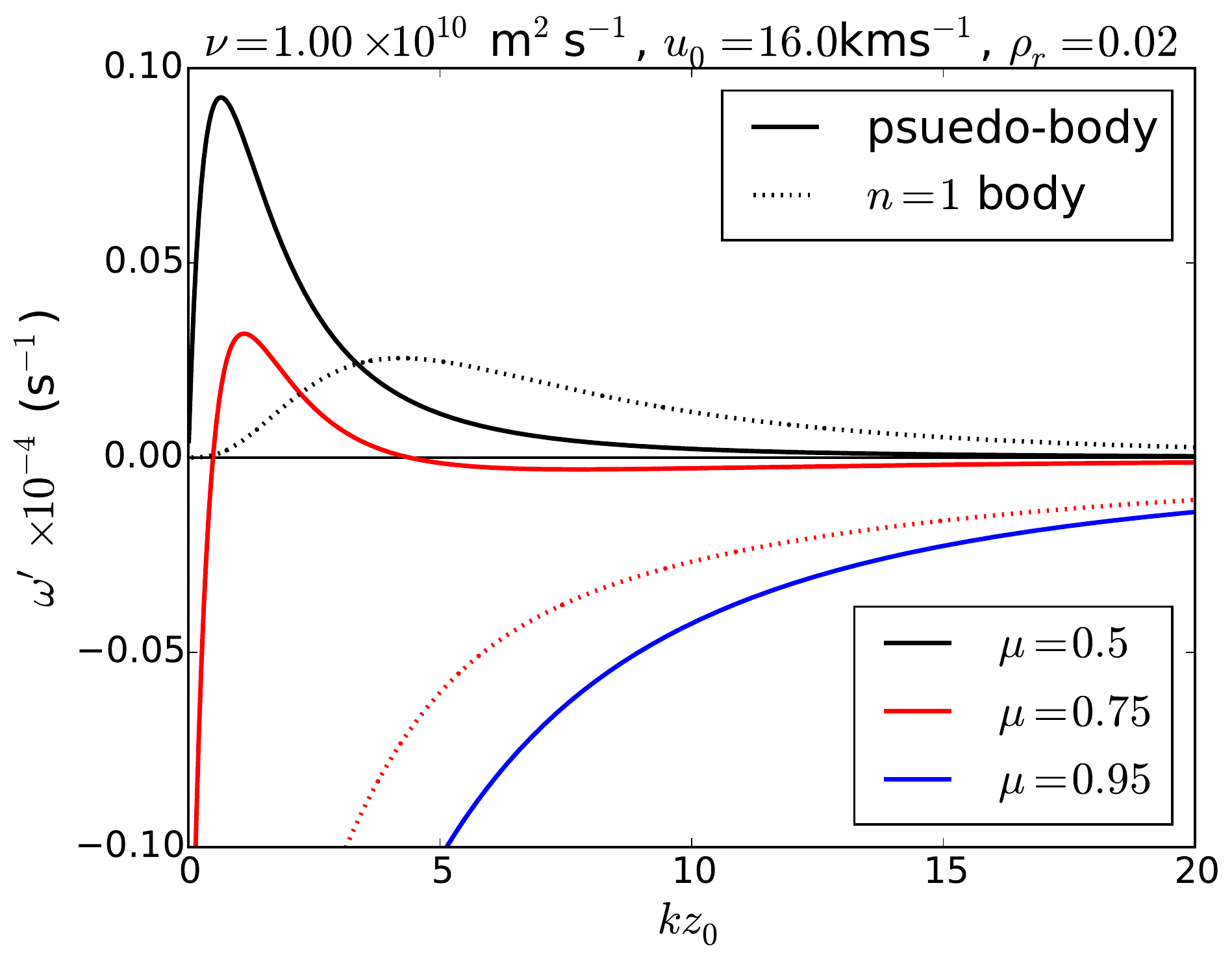}
		\caption{}
		\label{changek0013}
			\end{subfigure}
	\begin{subfigure}[b]{0.5\textwidth}
		\includegraphics[height=7cm,width=\textwidth]{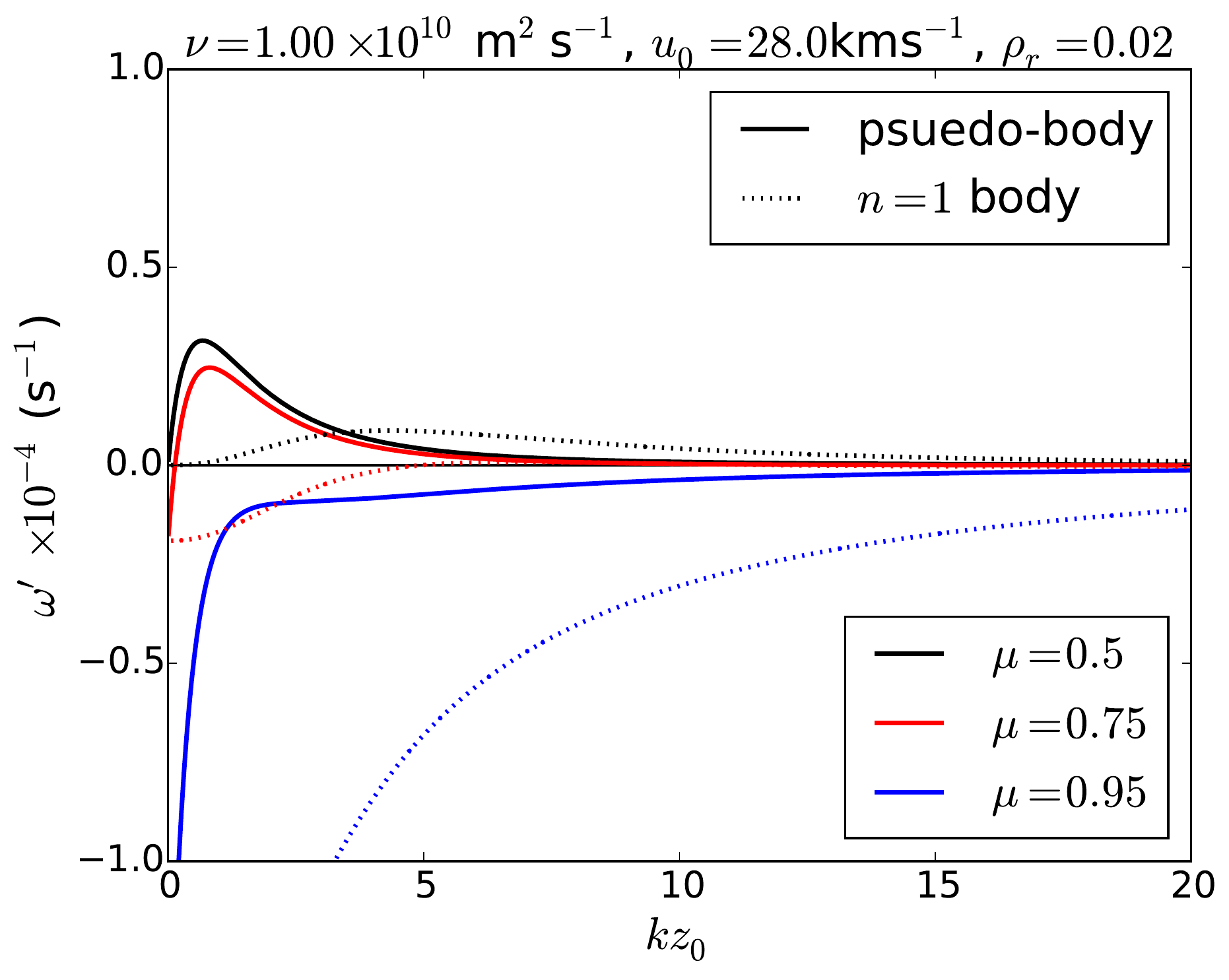} 	\caption{}
		\label{changek0014}
	\end{subfigure}
	\caption{Variation of the imaginary part of the frequency ($\omega'$) for pseudo-body (---) and $n=1$ ($\cdot$ $\cdot$) body mode with respect to dimensionless wave-number, $kz_0$. We fix the wavenumber at $k=5\times10^{-6}$ m${}^{-1}$ with the slabwidth varying. For each separate diagram the background flow speed has been taken to be $u_0=16.0$ km s${}^{-1}$ and $u_0=28.0$ km s${}^{-1}$, respectively. Panels (a) and (b) correspond to $\rho_r=0.01$, while panels (c) and (d) to $\rho_r=0.02$. All panels (a)-(d) use a viscosity coefficient of $\nu=10^{10}$ m${}^2$s${}^{-1}$. Black lines denote an ionisation degree of $\mu=0.5$, red and blue lines correspond to $\mu=0.75$ and $\mu=0.95$, respectively.}
	\label{changek001}
\end{figure*}
\begin{figure*}
	\begin{subfigure}[b]{0.5\textwidth}
		\includegraphics[height=7cm,width=\textwidth]{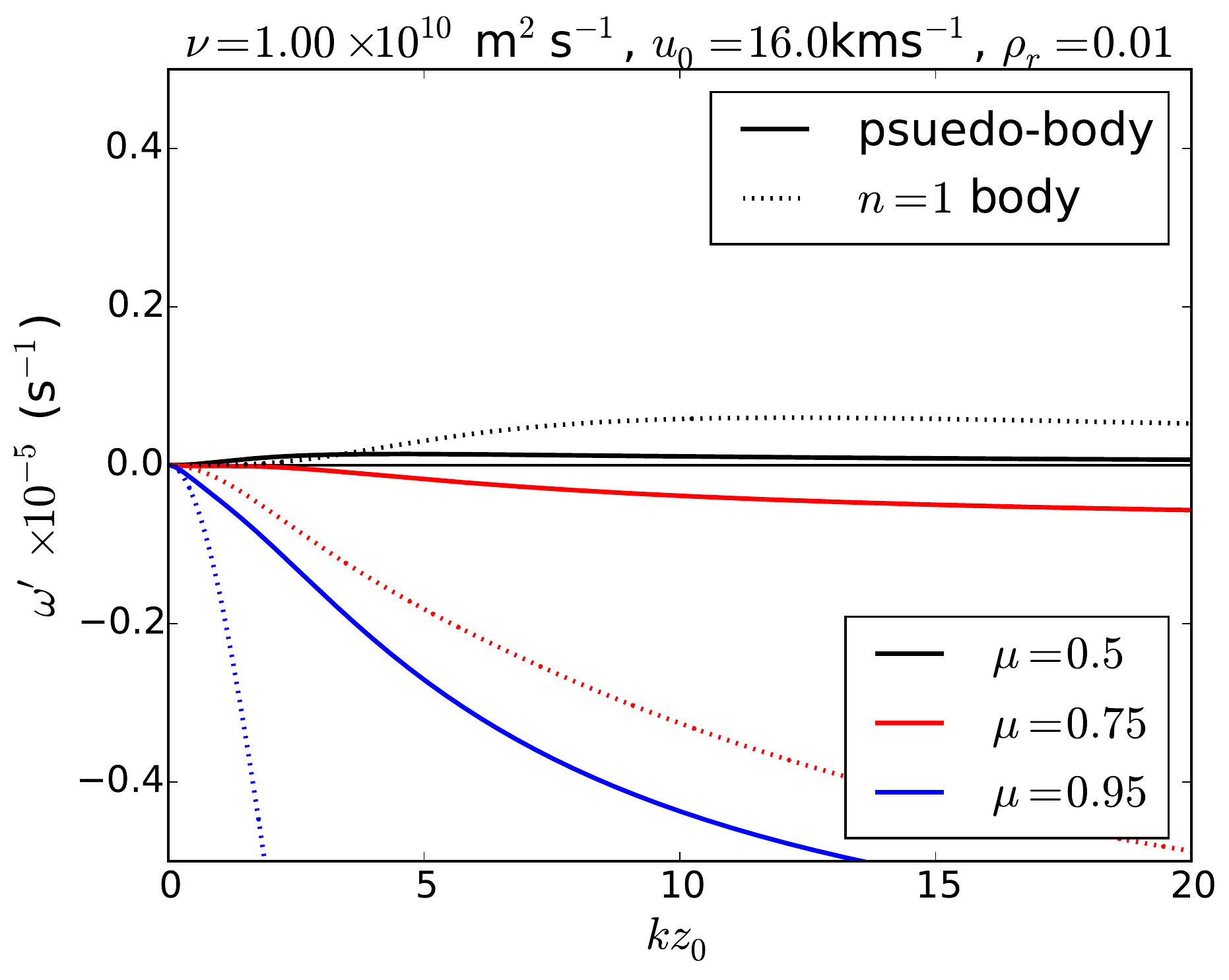}
		\caption{}
		\label{changek0021}
			\end{subfigure}
	\begin{subfigure}[b]{0.5\textwidth}
		\includegraphics[height=7cm,width=\textwidth]{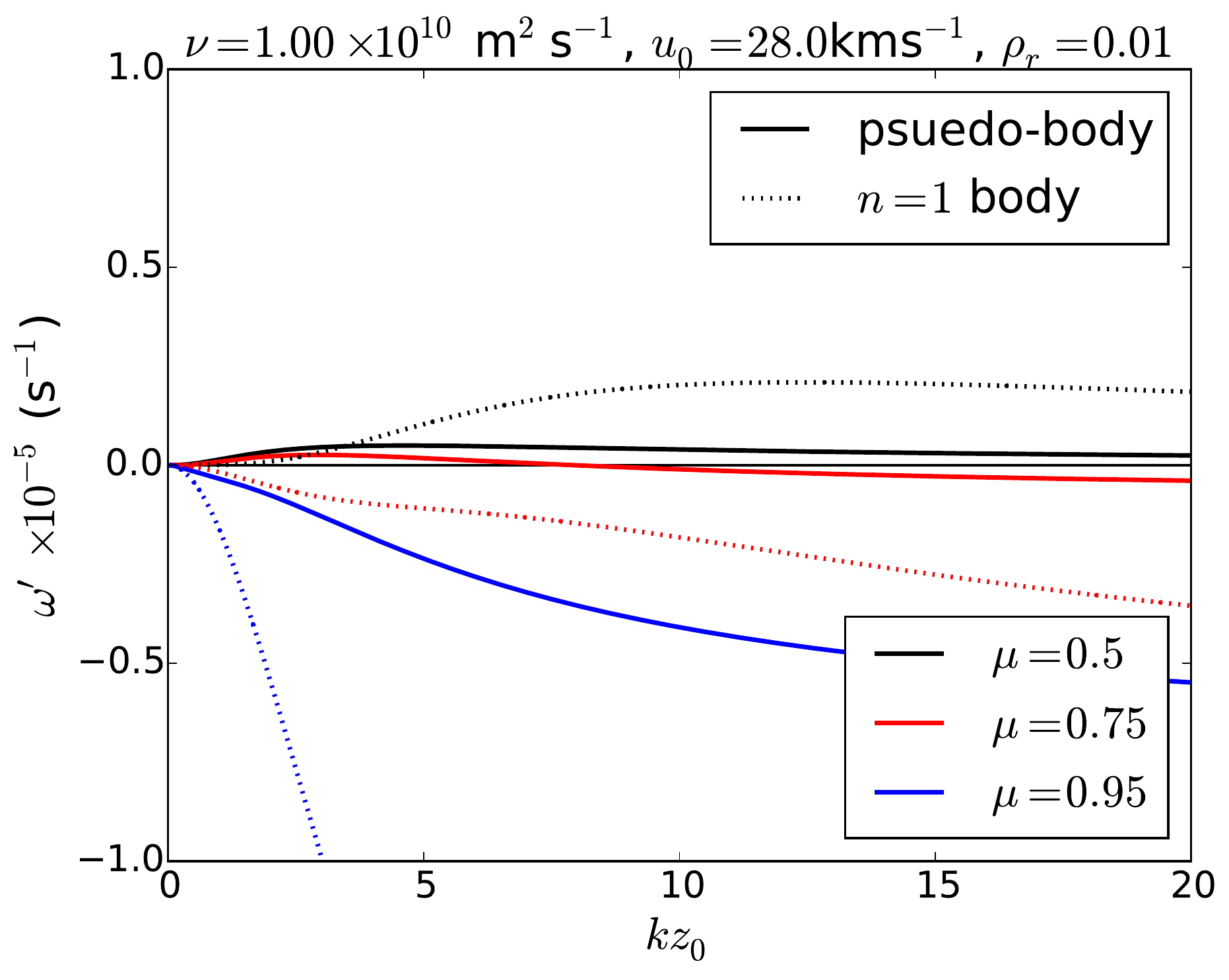} 	\caption{}
		\label{changek0022}
	\end{subfigure}
\begin{subfigure}[b]{0.5\textwidth}
		\includegraphics[height=7cm,width=\textwidth]{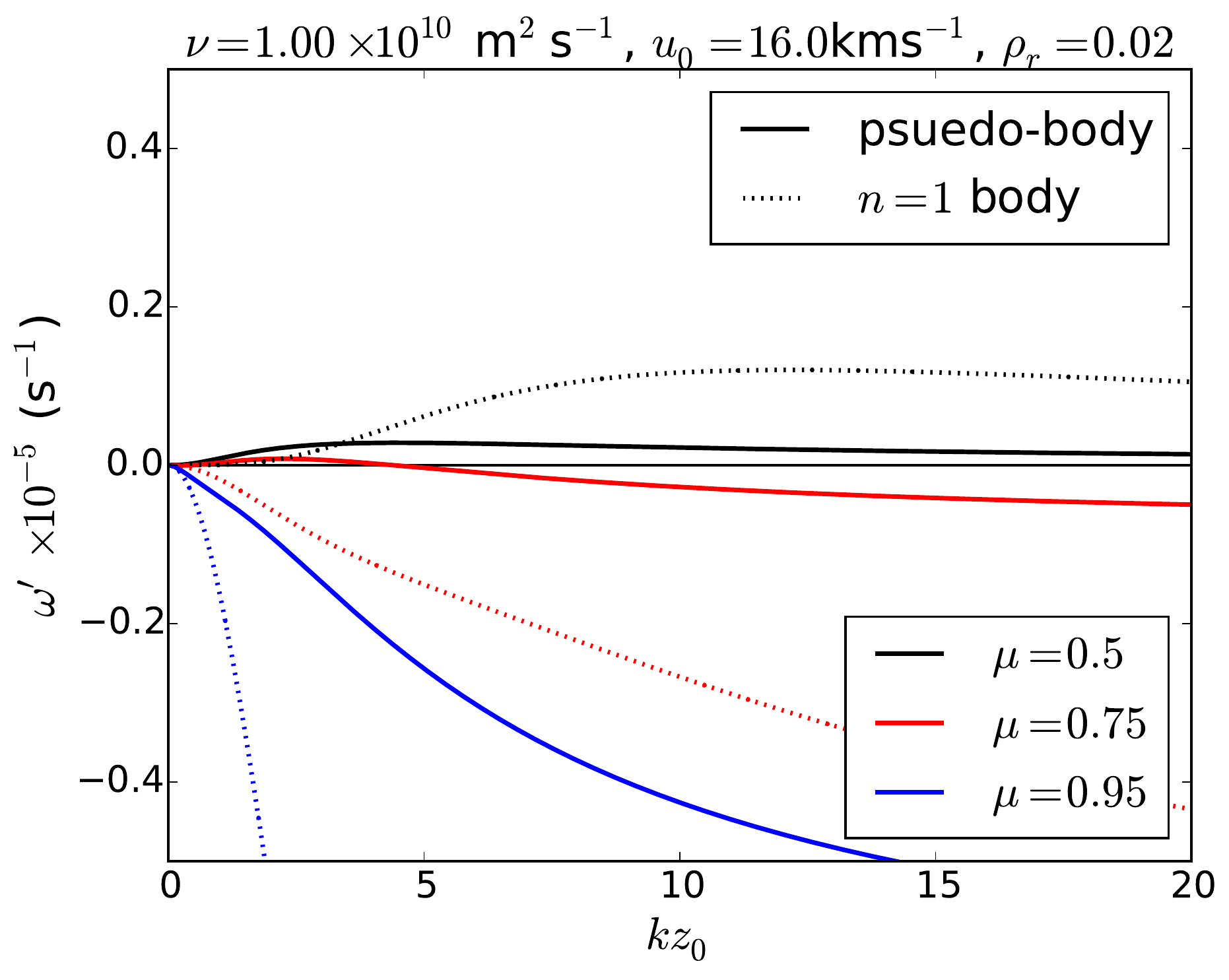}
		\caption{}
		\label{changek0023}
			\end{subfigure}
	\begin{subfigure}[b]{0.5\textwidth}
		\includegraphics[height=7cm,width=\textwidth]{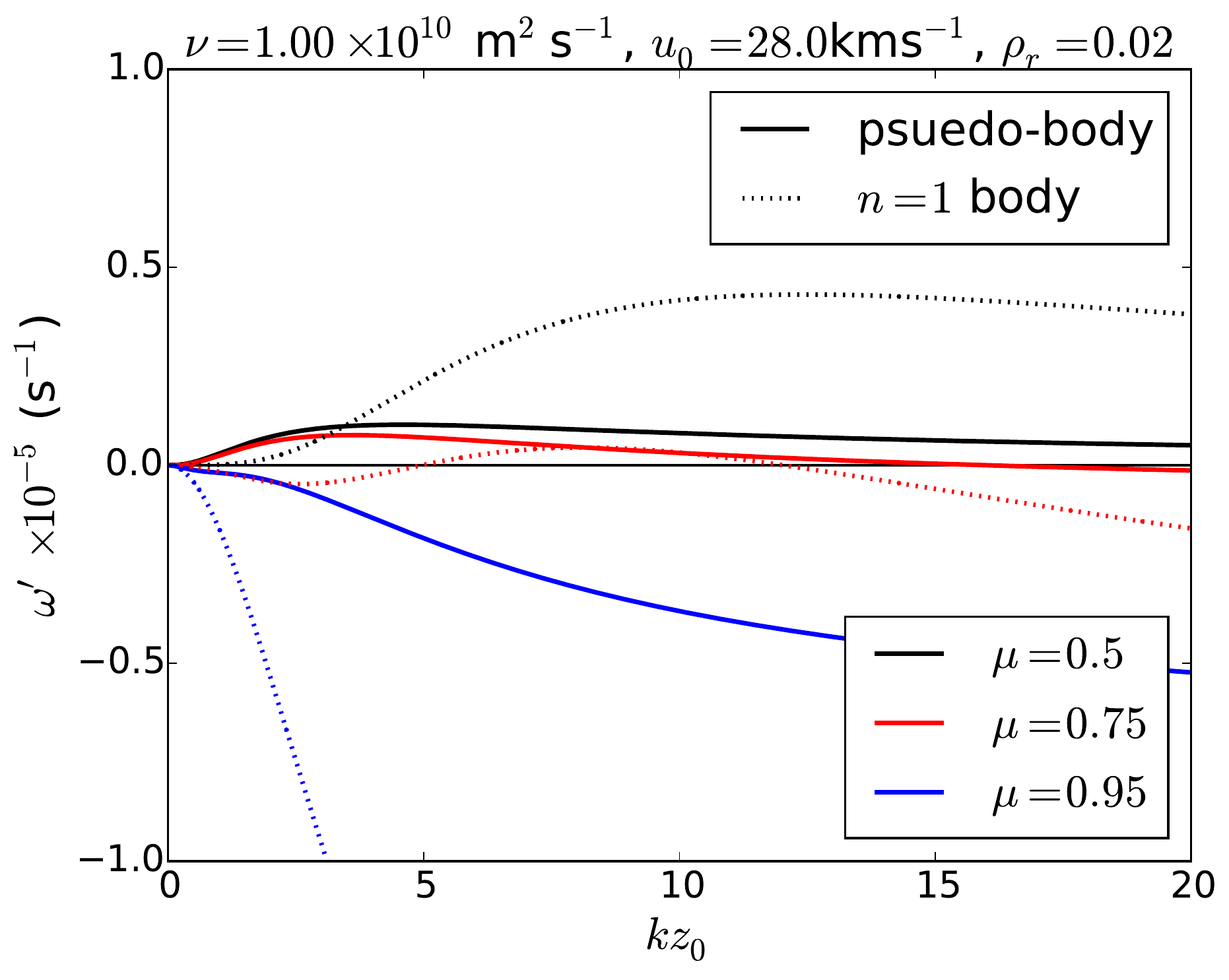} 	\caption{}
		\label{changek0024}
	\end{subfigure}
	\caption{Same as Fig.~\ref{changek001}, but with the width of the slab fixed at $z_0=4$ Mm and the wavenumber now the varying quantity.}
	\label{changek002}
\end{figure*}
For the case of a relatively cold chromospheric slab surrounded by hot coronal material with similar magnetic fields the density ratio is small i.e $\rho_r\ll1$. All the characteristic background speeds within the slab are much smaller than outside the slab. Since both plasma-$\beta$ values are small, the flows for which we have oscillatory solutions satisfy $u_0<c_{Te}\mp c_{T0}$.

In this limit, the imaginary part of the frequency given by Eq. (\ref{imaginarypart}) can be approximated as
	\begin{align}
		\omega'\approx\mp \dfrac{c_{s0}^4k^2\left[\dfrac{3}{2}\rho_r\nu(\pm c_{T0}+u_0)kz_0\pm \dfrac{\eta_{C}v_{A0}^2}{c_{T0}}\right]}{2c_{T0}\left(v_{A0}^2+c_{s0}^2\right)^2},
\label{imagfreqlimit}
	\end{align}
	where the two signs denote the forward and backward propagating wave. The straightfoward result that is obvious from Eq. \eqref{imagfreqlimit} is that $\omega'$ for the forward propagating wave (corresponding to the upper sign) is always negative, and, therefore, the forward propagating wave is always subject to damping. In contrast, the backward propagating wave (corresponding to the lower sign) is damped only for a particular combination of values, but $\omega'$ may become positive for flows that are larger than the critical value
	\begin{align}
		u_{0c}=c_{T0}+\dfrac{2\eta_{C}v_{A0}^2}{3\nu\rho_rkz_0	c_{T0}}.
\label{critflow}
	\end{align}  
In the absence of Cowling resitivity ($\eta_C=0$, i.e. the plasma is fully ionised), the instability occurs at flows speeds equal to $c_{T0}$. When neutrals are present in the system (i.e. $\eta_C\neq 0$) and the plasma is less ionised, the Cowling resistivity tends to stabilise the system by increasing the threshold where instability can appear. In the limit of an incompressible plasma, (i.e. $c_{s0}\to\infty$), the critical flow speed becomes
  	\begin{align}
		u_{0c}=v_{A0}+\dfrac{2\eta_{C}v_{A0}}{3\nu\rho_rkz_0},
\label{critflowincomp}
\end{align}
and, therefore, modes might become unstable for super-Alfv\'enic flows. Previous studies, for example, by \citet{2017A&A...603A..78B}, pointed out that in the case of sausage modes (see their Fig. 3), the necessary flow to have a dissipative instability was in the region of 30 km s${}^{-1}$. Using the same set of parameters, our Eq. \eqref{critflowincomp} results in a critical flow speed of only 28 km s${}^{-1}$, that is the two studies result in consistent critical values. The difference between the two values of critical flows may arise due to the fact that the value obtained by \citet{2017A&A...603A..78B} was obtained using numerical analysis, while the value obtained in the present paper is determined analytically. In order to obtain Eq. \eqref{critflowincomp}, several approximations were used (e.g. slender slab limit, dominant balance method), that could account for the small discrepancy between the two values.

It is interesting that in the case of an instability at a single interface the critical value obtained by \citet{2015A&A...577A..82B} is in the region of 48 km s${}^{-1}$. However, this value cannot be compared physically with our value, since while our discussion is valid in the thin slab approximation (i.e. $kz_0\to 0$), the results of \citet{2015A&A...577A..82B} are valid provided $kz_0\to \infty$. However, we may conclude that the dissipative instability in a magnetic slab requires a lower critical flow speed for its development. Therefore the result given in Eq. \eqref{critflow} suggests that once compressibility is taken into account, the plasma can become unstable at lower flow speeds ($c_{T0}<v_{A0}$) (a lower critical flow speed is needed for the dissipative instability to occur), so compressibility tends to destabilise the plasma. 

The whole partially ionised prominence is modelled here by a slab that is surrounded by the fully ionised corona as specified earlier. In general, the width of prominences varies between around 1-30 Mm (\citealp{2011SSRv..158..237L}). Observations of waves in prominences show that typical wave-numbers are between $10^{-8}$ and $10^{-6}$ m${}^{-1}$, meaning that the slender slab limit is justified to a large extent, however, this does not cover the whole spectrum of possible values. 

 We consider a slab width of \mbox{$\approx1$ Mm}, so that the set value of \mbox{$kz_0=0.01$} corresponds to a wave-number of \mbox{$k=2\times10^{-7}$ m${}^{-1}$}, while the dimensionless quantity  \mbox{$kz_0=0.02$}, corresponds to the  wave-number \mbox{$k=4\times10^{-7}$ m${}^{-1}$}.
We assume that the temperature of the prominence is \mbox{$T_0=10^4$ K} that corresponds to a sound speed speed of \mbox{$c_{s0}=11.7$ km s${}^{-1}$}. Assuming an Alfv\'en speed of \mbox{{$v_{A0}=28.0$ {km s}${}^{-1}$}} results in a tube speed of \mbox{$c_{T0}=10.8$ km s${}^{-1}$}. When $\rho_r=0.02$ the coronal characteristic speed are (assuming the same magnetic field strength in all regions):
\mbox{$v_{Ae}=198.0$ km s${}^{-1}$}, \mbox{$c_{se}=83.0$ km s${}^{-1}$} and \mbox{$c_{Te}=76.5$ km s${}^{-1}$}. When $\rho_r=0.01$ the coronal characteristic speeds become \mbox{$v_{Ae}=280.0$ km s${}^{-1}$}, \mbox{$c_{se}=117.0$ km s${}^{-1}$} and \mbox{$c_{Te}=108.0$ km s${}^{-1}$}.

The variation of the critical flow speed with the ionisation degree for compressible (based on Eq. \ref{critflow}) and incompressible (based on Eq. \ref{critflowincomp}) plasma is shown in Figure \ref{compression}. The top two panels correspond to a viscosity coefficient of \mbox{$\nu=10^{9}$ m${}^2$s${}^{-1}$} that leads to a viscous Reynolds number of $R\approx10^4$ (this may be small, however it has been shown in \cite{1985JGR....90.7620H} that in particular coronal conditions a small Reynolds number can be valid) with typical length scales of $10^7$ m (again, a typically observed wavelength) and typical background wave velocities of $10^5$-$10^6$ m s${}^{-1}$, that is the external Alfv\'en and sound speeds. The variation of the critical speed is studied for two distinct values of the density ratio between corona and chromosphere ($\rho_r=0.01$ and $\rho_r=0.02$). The bottom two panels show the variation of the critical flow speed for a viscosity coefficient of $\nu=10^{10}$ m${}^2$s${}^{-1}$ (corresponding to a Reynolds number of $R\approx10^3$), for the same density ratios as before. In these plots, the ionisation degree, $\mu$, varies between 0.5 (fully ionised plasma) to one (fully neutral fluid) and for different values of the dimensionless quantity $kz_0$. 

These figures clearly reveal that, in the case of a fully ionised plasma, the instability occurs at flows speeds that have realistic values (see e.g. \citealp{2005SoPh..226..239L}, \citealp{2008ASPC..383..235L}) and the critical value of the flow for compressional plasma is approximately half of the flow necessary to induce an instability in an incompressible plasma. As the concentration of neutrals is increased, there is a critical value of this ionisation degree after which the critical flow corresponding to the incompressible plasma is lower. At this point, the curves representing the critical flows intersect. When either $kz_0$ or $\rho_r$ are increased the value of $\mu$ at which the solutions cross is increased and the gradients of the flow with respect to $\mu$ decrease. It is possible to find the critical value, $\mu_c$, when the solution paths cross, by equating Eqs. \eqref{critflow} and \eqref{critflowincomp}
\begin{equation}
\mu_c=\dfrac{1+\dfrac{2\hat{\eta}_C}{3\nu\rho_rkz_0}\dfrac{v_{A0}}{c_{T0}}}{1+\dfrac{4\hat{\eta}_C}{3\nu\rho_rkz_0}\dfrac{v_{A0}}{c_{T0}}}.
\label{critmu}
\end{equation}

The role of various physical parameters in the appearance of the dissipative instability can be deduced from Fig. 2. First of all, we keep in mind that the observed flows are sub-sonic.  Comparing the two left-hand panels, we see that, with a more viscous corona, a realistic range of flow speeds for the instability to occur can be obtained even with a large fraction of neutrals present. In the top left-hand side panel, the realistic critical flows could appear only for plasmas that are nearly completely ionised, however, with the increase of viscosity, the plasma is more prone to instability, even for an increased number of neutrals. That means that viscosity has a destabilising effect. In all panels it is obvious that once the concentration of neutrals is increased the critical speed where instability occurs increases, meaning that neutrals have a stabilising effect. Comparing the upper two panels we can also observe that larger wavelength waves are more easily developing instability. For a given ionisation degree waves with a larger wavenumber (shorter wavelength) become unstable. 

Figure \ref{imagpart} plots the variation of the imaginary part of the frequency, $\omega'$, with respect to the ionisation degree ($\mu$) and the internal equilibrium flow speed for $\rho_r=0.01$ and $kz_0=0.1$ (Figures \ref{imagpart1} and \ref{imagpart2}) and $\rho_r=0.02$ and $kz_0=0.2$ (Figures \ref{imagpart3} and \ref{imagpart4}). The dividing line, labelled `$\omega'=0$', shows the transition between damped (stable) waves ($\omega'<0$) and amplified (unstable) waves ($\omega'>0$). 

 In Figs \ref{imagpart1} and \ref{imagpart3} the wave-number and viscosity coefficient take the values of \mbox{$k=2\times10^{-7}$ m${}^{-1}$} and \mbox{$\nu=10^{9}$ m${}^2$ s${}^{-1}$}, while in Figs. \ref{imagpart2} and \ref{imagpart4} we used \mbox{$k=4\times10^{-7}$ m${}^{-1}$} and \mbox{$\nu=10^{10}$ m${}^2$ s${}^{-1}$}. It is evident that for a given ionisation degree, as $u_0$ increases, damping is reduced below the contour $\omega'=0$ s${}^{-1}$ and that amplification increases above this. 

As the fraction of neutrals increases the magnitude of the damping becomes larger below the $\omega'=0$ s${}^{-1}$ contour and the magnitude of the amplification decreases above this. For a given equilibrium flow strength above the critical speed described by Eq. (\ref{critflow}), there will be a critical ionisation degree after which the plasma becomes stable, meaning that the presence of neutrals, therefore, acts to stabilise the prominence. As the coefficient of viscosity of the corona is increased, the parameter domain where instability can arise is much more extended.

The damping and amplification time-scales in Figs. \ref{imagpart1} and \ref{imagpart3} are larger than $10^{7}$ s which is clearly far too long for any active prominence structure. 

\subsection{Numerical solutions}

The analysis using the slender slab limit is very useful as a guide towards understanding the nature and behaviour of the modes present in the system. However, this limit is restrictive due to the size of prominences, for which the condition $kz_0\ll1$ is not satisfied. Therefore, it is instructive to solve numerically Eq. \eqref{Sausgeideal} and approximate the imaginary part of the frequency using Eq. \eqref{sausageimag}.

Figures \ref{changek001} and \ref{changek002} show the variation of the imaginary part of the frequency ($\omega'$), with respect to the dimensionless wave-number, $kz_0$, for three different values of the ionisation degree and the same viscosity coefficient. In both figures, the top panels correspond to a density ratio of \mbox{$\rho_r=0.01$}, while the bottom panels were obtained for \mbox{$\rho_r=0.02$}. The value of the background flow was chosen to be \mbox{$u_0=16$ km s${}^{-1}$} (just above the internal tube speed, $c_{T0}$, in the left-hand side panels) and \mbox{$u_0=28$ km s${}^{-1}$} (close to the internal Alfv\'en speed, $v_{A0}$, in right-hand side panels), respectively.  The variation of the dimensionless damping or amplification rate is plotted against the dimensionless quantity $kz_0$, keeping $k=5\times 10^{-6}$ m$^{-1}$ and allowing the width of the slab, $2z_0$, to vary (see Fig. \ref{changek001}). In Fig. \ref{changek002}, we plot the same quantity, but now the width of the slab is maintained constant at $z_0=4$ Mm (Fig. \ref{changek001}) and $k$ is allowed to vary. In all cases a horizontal is drawn at the $\omega'=0$ level in order to clearly identify the behaviour of the rate of change of the amplitude.

Figure \ref{changek001} shows that at flow speeds just above the internal tube speed of the prominence, the pseudo-body mode and the \mbox{$n=1$} body modes are unstable when $\mu=0.5$, that is when the prominence is fully ionised. As the ionisation fraction increases (that is more neutrals are taken into account) all modes become stable and their amplitudes are damped, with the pseudo-body most being the more damped. This results confirms earlier findings that neutrals have a stabilising effect. As $kz_0$ is increased (larger slab sizes) both the damping and amplification of modes is reduced. When the equilibrium flow is increased to \mbox{$u_0=28$ km s${}^{-1}$} (Fig. \ref{changek0012}) modes will use the increased flow speed for additional energy, waves will become unstable much easier. Now, the increased amount of neutrals is not enough to stabilise completely the body mode; the body mode corresponding to $\mu=0.75$ is stable only for very a thin slab. For an even larger concentration of neutrals ($\mu=0.95$), both the body and pseudo-body mode are damped. Comparing the results obtained for the two flow values we can observe that the amplification rate increases with the value of the equilibrium flow. 

When the density contrast between the prominence and corona is increased the qualitative behaviour of modes in the fully ionised case does not change, however, significant variations occur in the behaviour of the pseudo mode corresponding to $\mu=0.75$. At this concentration of neutrals, the pseudo-body mode remains stable only for large wavelengths after which it becomes unstable. For a value of $kz_0\approx 4$, the modes becomes stable again. This latter effect could be attributed to dispersion. 
Finally, when the equilibrium flow is increased the pseudo-body mode corresponding to an ionisation degree of 0.75 stays stable only for very large wavelengths. The $n=1$ body modes becomes unstable for a given value of $kz_0$. These results confirms our finding that an increased flow speed is generating instability in modes.

When the magnetic slab width is kept constant (Fig. \ref{changek002}) the variation of the imaginary part of frequency is investigated between the limits of long and short wavelength. All figures reveal that the most unstable or damped regime appears in the long wavelength approximation, while for small wavelengths all rates tend to zero. Similar to the previous case, the amplification or damping rate increases by almost an order of magnitude when the flow speed is increased. The maximum of the amplification rate occurs at larger wavelengths, once the concentration of neutrals is increased. Similar enhancement of the damping or amplification rate can be observed when the density contrast is increased.

\section{Conclusions}

The main focus of the current investigation is the dissipative instability associated with negative energy waves (\citealp{1988ZhETF..94..138R}), that is backward propagating waves that are amplified during their propagation in a partially ionised prominence plasma slab surrounded by a fully ionised and viscous corona. We derived the dispersion relation for compressible sausage modes (see Eq. \ref{sausagedisp}), described by a highly transcendental equation. The frequency of modes is approximated using a regular perturbation method in the slender slab limit, that is when $kz_0\ll 1$, and two solutions were found: one mode that propagates with the phase speed of the external sound speed and the other mode that progresses at the phase speed of the internal tube speed, $c_{T0}$, Doppler-shifted with the internal flow speed, $u_0$. For the second of these two cases, the phase speed of the backward propagating mode reverses direction when the internal flow speed is approximately equal to the internal tube speed. Our study deals only with the stability question of slow sausage modes, as these are the most likely to become unstable given the thresholds of equilibrium flow speeds.

The coupling of backward propagating modes and the equilibrium flow may result in reversing the propagation direction of modes. Further, the flow serves as a source of energy for wave amplification. The amplification rate is proportional to the value of dissipative coefficients. The instability only occurs for flow speeds larger than the internal tube speed and this threshold increases when more neutrals are present in the system. In contrast, the critical speed is inversely proportional to the viscosity coefficient of the fully ionised corona. In addition, the critical speed increases with the increase in the wavelength. This result is rather remarkable, as it shows that modes propagating along the prominence slab can become unstable for flows that are in the observable range (in contrast, a Kelvin-Helmholtz instability requires flow speeds that are super-Alfv\'enic). 

To investigate the effect of compressibility we compared our results to the results obtained in the incompressible limit, a similar case that is found in \citet{2017A&A...603A..78B} for long wavelength limit. First of all, the critical flow threshold derived by us agrees very well with the values obtained for sausage modes by \citet{2017A&A...603A..78B}, but these are both super Alfv\'enic flows. For a compressible plasma, the critical speed was obtained to be sub-Alfv\'enic, which is more realistic, as it lies in the mid-range of values for equilibrium flows presently observed in solar prominences. The flow values determined by us are much smaller than the critical values determined by \citet{2015A&A...577A..82B} for the single interface, meaning that guided waves need smaller flow speed to become unstable. However, depending on the ionisation degree of the plasma, this statement does not hold for a larger proportion of neutrals. Therefore, when the plasma has a sufficient amount of neutrals present, the interface is more stable than in the incompressible case.  It is worth comparing the main results of the present study with the results by  \citet{2017A&A...603A..78B}. While the value of the characteristic speed for which backward propagating waves in the slender slab limit ($kz_0=0.1$ for comparison purposes) become unstable was found to be in the region of 30 km s$^{-1}$, our analysis shows that these waves, in the presence of compression, can become unstable for speeds of the order of 10 km s$^{-1}$. Inspecting the value of the imaginary part of the frequency (the quantity that describe the instability increment of waves), the values obtained in the compressional limit would be at least three orders of magnitude smaller than the values obtained by  \citet{2017A&A...603A..78B}. This reveals the fact that the compressibility can also have a stabilising effect on the plasma.   

The imaginary part of the frequency varied with flow speed and ionisation degree and in all investigated case these instabilities proved to be slow, the time scales involved could well be of the order of $10^7$ seconds in the long wavelength limit. The numerical analysis of the dispersion relation for a large spectrum of parameters reveals that the amplification rate can reach values of the order of $10^{-5}$ s$^{-1}$ such that instabilities have a characteristic time of $10^5$ seconds

The slender slab limit, while being fairly instructive for analytically modelling the properties of the modes and instabilities, did not provide a full picture of the behaviour of modes and the stability of the plasma structures. That is why we solved the full dispersion relation numerically in order to investigate how varying the importance of dispersive effects change the imaginary party of the frequency. For pseudo-body modes the largest amplification rate was obtained for a flow speed of 28 km s${}^{-1}$ and the maximum value was attained for $kz_0\approx 1$, that is a wavelength which is comparable with the transversal size of the slab. The amplification rate of the wave amplitudes dropped off rapidly with increasing $kz_0$. Again, the absolute value of the amplification rate and damping rate was fairly of course very dependent to changes in $\nu$ and $\rho_r$. For the body-mode ($n=1$), the effect of a larger proportion of neutrals is to damp the waves more significantly than the previous body mode. The amplification for these modes, caused by the dissipative instability, was much lower than for the pseudo-body mode, so they are less likely to cause unstable behaviour than the pseudo-body mode would.

In conclusion, our numerical analysis reveals that for a given density ratio and flow speed the inclusion of neutrals stabilises the plasma. This result could be easily understood, however, it difficult to prove given the framework used to describe the instability. Once the number of neutrals increases, the collisions between heavy particles (ions and neutrals) will distribute the energy stored in waves. However, this aspect cannot be evidenced in a single fluid description. Our aim is to expand in the near future the current analysis to a two-fluid model and investigate the role of collisions between heavy particles in the stability of the plasma.

Increasing the value of the flow will increase the damping or amplification rate of modes making the stable modes become unstable, therefore the flow has a destabilising effect. This conclusion is again easy to understand keeping in mind that the dissipative instability (and implicitly negative energy waves) use the background equilibrium flow as a source of energy. The amplification of waves is more pronounced for short wavelengths and intermediate slab thickness.

\section*{Acknowledgements}
The present research was supported by the Leverhulme Trust (IN-2014-016). The authors are grateful to the Science and Technology Facilities Council (STFC) UK and The Royal Society, UK. I.B. was partly supported by a grant of the Ministry of National Education and Scientific Research, RDI Programme for Space Technology and Advanced Research - STAR, project number 181/20.07.2017.

\bibliographystyle{aa}
\bibliography{paper}       

\begin{thebibliography}{40}
\expandafter\ifx\csname natexlab\endcsname\relax\def\natexlab#1{#1}\fi

\bibitem[{{Arregui} {et~al.}(2012){Arregui}, {Oliver}, \&
  {Ballester}}]{2012LRSP....9....2A}
{Arregui}, I., {Oliver}, R., \& {Ballester}, J.~L. 2012, Living Reviews in
  Solar Physics, 9, 2

\bibitem[{{Ballai} {et~al.}(2015){Ballai}, {Oliver}, \&
  {Alexandrou}}]{2015A&A...577A..82B}
{Ballai}, I., {Oliver}, R., \& {Alexandrou}, M. 2015, \aap, 577, A82

\bibitem[{{Ballai} {et~al.}(2017){Ballai}, {Pint{\'e}r}, {Oliver}, \&
  {Alexandrou}}]{2017A&A...603A..78B}
{Ballai}, I., {Pint{\'e}r}, B., {Oliver}, R., \& {Alexandrou}, M. 2017, \aap,
  603, A78

\bibitem[{{Berger} {et~al.}(2008){Berger}, {Shine}, {Slater}, {Tarbell},
  {Title}, {Okamoto}, {Ichimoto}, {Katsukawa}, {Suematsu}, {Tsuneta}, {Lites},
  \& {Shimizu}}]{2008ApJ...676L..89B}
{Berger}, T.~E., {Shine}, R.~A., {Slater}, G.~L., {et~al.} 2008, \apjl, 676,
  L89

\bibitem[{{Berger} {et~al.}(2010){Berger}, {Slater}, {Hurlburt}, {Shine},
  {Tarbell}, {Title}, {Lites}, {Okamoto}, {Ichimoto}, {Katsukawa}, {Magara},
  {Suematsu}, \& {Shimizu}}]{2010ApJ...716.1288B}
{Berger}, T.~E., {Slater}, G., {Hurlburt}, N., {et~al.} 2010, \apj, 716, 1288

\bibitem[{{Braginskii}(1965)}]{1965RvPP....1..205B}
{Braginskii}, S.~I. 1965, Reviews of Plasma Physics, 1, 205

\bibitem[{{Cairns}(1979)}]{1979JFM....92....1C}
{Cairns}, R.~A. 1979, J. Fluid Mech., 92, 1

\bibitem[{{Edwin} \& {Roberts}(1982)}]{1982SoPh...76..239E}
{Edwin}, P.~M. \& {Roberts}, B. 1982, \solphys, 76, 239

\bibitem[{{Engvold}(1998)}]{1998ASPC..150...23E}
{Engvold}, O. 1998, in Astronomical Society of the Pacific Conference Series,
  Vol. 150, IAU Colloq. 167: New Perspectives on Solar Prominences, ed. D.~F.
  {Webb}, B.~{Schmieder}, \& D.~M. {Rust}, 23

\bibitem[{{Erd{\'e}lyi} \& {Fedun}(2006)}]{2006SoPh..238...41E}
{Erd{\'e}lyi}, R. \& {Fedun}, V. 2006, \solphys, 238, 41

\bibitem[{{Erd{\'e}lyi} \& {Fedun}(2007)}]{2007SoPh..246..101E}
{Erd{\'e}lyi}, R. \& {Fedun}, V. 2007, \solphys, 246, 101

\bibitem[{{Heinzel} \& {Anzer}(2006)}]{2006ApJ...643L..65H}
{Heinzel}, P. \& {Anzer}, U. 2006, \apjl, 643, L65

\bibitem[{{Hillier} {et~al.}(2012){Hillier}, {Berger}, {Isobe}, \&
  {Shibata}}]{2012ApJ...746..120H}
{Hillier}, A., {Berger}, T., {Isobe}, H., \& {Shibata}, K. 2012, \apj, 746, 120

\bibitem[{{Hollweg}(1985)}]{1985JGR....90.7620H}
{Hollweg}, J.~V. 1985, \jgr, 90, 7620

\bibitem[{{Joarder} {et~al.}(1997){Joarder}, {Nakariakov}, \&
  {Roberts}}]{1997SoPh..176..285J}
{Joarder}, P.~S., {Nakariakov}, V.~M., \& {Roberts}, B. 1997, \solphys, 176,
  285

\bibitem[{{Khodachenko} {et~al.}(2004){Khodachenko}, {Arber}, {Rucker}, \&
  {Hanslmeier}}]{2004A&A...422.1073K}
{Khodachenko}, M.~L., {Arber}, T.~D., {Rucker}, H.~O., \& {Hanslmeier}, A.
  2004, \aap, 422, 1073

\bibitem[{{Lin} {et~al.}(2005{\natexlab{a}}){Lin}, {Banerjee}, {Doyle}, \&
  {O'Shea}}]{2005A&A...444..585L}
{Lin}, C.-H., {Banerjee}, D., {Doyle}, J.~G., \& {O'Shea}, E.
  2005{\natexlab{a}}, \aap, 444, 585

\bibitem[{{Lin}(2011)}]{2011SSRv..158..237L}
{Lin}, Y. 2011, \ssr, 158, 237

\bibitem[{{Lin} {et~al.}(2005{\natexlab{b}}){Lin}, {Engvold}, {Rouppe van der
  Voort}, {Wiik}, \& {Berger}}]{2005SoPh..226..239L}
{Lin}, Y., {Engvold}, O., {Rouppe van der Voort}, L., {Wiik}, J.~E., \&
  {Berger}, T.~E. 2005{\natexlab{b}}, \solphys, 226, 239

\bibitem[{{Lin} {et~al.}(2008){Lin}, {Martin}, \&
  {Engvold}}]{2008ASPC..383..235L}
{Lin}, Y., {Martin}, S.~F., \& {Engvold}, O. 2008, in Astronomical Society of
  the Pacific Conference Series, Vol. 383, Subsurface and Atmospheric
  Influences on Solar Activity, ed. R.~{Howe}, R.~W. {Komm}, K.~S.
  {Balasubramaniam}, \& G.~J.~D. {Petrie}, 235

\bibitem[{{Malville} \& {Schindler}(1981)}]{1981SoPh...70..115M}
{Malville}, J.~M. \& {Schindler}, M. 1981, \solphys, 70, 115

\bibitem[{{Molowny-Horas} {et~al.}(1997){Molowny-Horas}, {Oliver}, {Ballester},
  \& {Baudin}}]{1997SoPh..172..181M}
{Molowny-Horas}, R., {Oliver}, R., {Ballester}, J.~L., \& {Baudin}, F. 1997,
  \solphys, 172, 181

\bibitem[{{Nakariakov} \& {Roberts}(1995)}]{1995SoPh..159..213N}
{Nakariakov}, V.~M. \& {Roberts}, B. 1995, \solphys, 159, 213

\bibitem[{{Okamoto} {et~al.}(2007){Okamoto}, {Tsuneta}, {Berger}, {Ichimoto},
  {Katsukawa}, {Lites}, {Nagata}, {Shibata}, {Shimizu}, {Shine}, {Suematsu},
  {Tarbell}, \& {Title}}]{2007Sci...318.1577O}
{Okamoto}, T.~J., {Tsuneta}, S., {Berger}, T.~E., {et~al.} 2007, Science, 318,
  1577

\bibitem[{{Parenti}(2014)}]{2014LRSP...11....1P}
{Parenti}, S. 2014, Living Reviews in Solar Physics, 11, 1

\bibitem[{Ruderman(2005)}]{doi:10.1063/1.1856931}
Ruderman, M.~S. 2005, Physics of Plasmas, 12, 034701

\bibitem[{{Ruderman} {et~al.}(1996){Ruderman}, {Verwichte}, {Erd{\'e}lyi}, \&
  {Goossens}}]{1996JPlPh..56..285R}
{Ruderman}, M.~S., {Verwichte}, E., {Erd{\'e}lyi}, R., \& {Goossens}, M. 1996,
  J. Plasma Phys., 56, 285

\bibitem[{{Ruderman} \& {Wright}(1998)}]{1998JGR...10326573R}
{Ruderman}, M.~S. \& {Wright}, A.~N. 1998, \jgr, 103, 26573

\bibitem[{{Ruzdjak} \& {Tandberg-Hanssen}(1990)}]{1990LNP...363.....R}
{Ruzdjak}, V. \& {Tandberg-Hanssen}, E., eds. 1990, Lecture Notes in Physics,
  Berlin Springer Verlag, Vol. 363, {Dynamics of quiescent prominences;
  Proceedings of the 117th Colloquium of IAU, Hvar, Yugoslavia, Sept. 25-29,
  1989}

\bibitem[{{Ryutova} {et~al.}(2010){Ryutova}, {Berger}, {Frank}, {Tarbell}, \&
  {Title}}]{2010SoPh..267...75R}
{Ryutova}, M., {Berger}, T., {Frank}, Z., {Tarbell}, T., \& {Title}, A. 2010,
  \solphys, 267, 75

\bibitem[{{Ryutova} {et~al.}(2012){Ryutova}, {Berger}, {Frank}, {Title}, \&
  {Tarbell}}]{2012ASPC..454..143R}
{Ryutova}, M., {Berger}, T., {Frank}, Z., {Title}, A., \& {Tarbell}, T. 2012,
  in Astronomical Society of the Pacific Conference Series, Vol. 454, Hinode-3:
  The 3rd Hinode Science Meeting, ed. T.~{Sekii}, T.~{Watanabe}, \&
  T.~{Sakurai}, 143

\bibitem[{{Ryutova}(1988)}]{1988ZhETF..94..138R}
{Ryutova}, M.~P. 1988, Zhurnal Eksperimentalnoi i Teoreticheskoi Fiziki, 94,
  138

\bibitem[{{Schmieder} {et~al.}(1984){Schmieder}, {Malherbe}, {Mein}, \&
  {Tandberg-Hanssen}}]{1984A&A...136...81S}
{Schmieder}, B., {Malherbe}, J.~M., {Mein}, P., \& {Tandberg-Hanssen}, E. 1984,
  \aap, 136, 81

\bibitem[{{Terra-Homem} {et~al.}(2003){Terra-Homem}, {Erd{\'e}lyi}, \&
  {Ballai}}]{2003SoPh..217..199T}
{Terra-Homem}, M., {Erd{\'e}lyi}, R., \& {Ballai}, I. 2003, \solphys, 217, 199

\bibitem[{{Terradas} {et~al.}(2002){Terradas}, {Molowny-Horas}, {Wiehr},
  {Balthasar}, {Oliver}, \& {Ballester}}]{2002A&A...393..637T}
{Terradas}, J., {Molowny-Horas}, R., {Wiehr}, E., {et~al.} 2002, \aap, 393, 637

\bibitem[{{Thompson} \& {Schmieder}(1991)}]{1991A&A...243..501T}
{Thompson}, W.~T. \& {Schmieder}, B. 1991, \aap, 243, 501

\bibitem[{{Tirry} {et~al.}(1998){Tirry}, {Cadez}, {Erdelyi}, \&
  {Goossens}}]{1998A&A...332..786T}
{Tirry}, W.~J., {Cadez}, V.~M., {Erdelyi}, R., \& {Goossens}, M. 1998, \aap,
  332, 786

\bibitem[{{Zaqarashvili} {et~al.}(2010){Zaqarashvili}, {D{\'{\i}}az}, {Oliver},
  \& {Ballester}}]{2010A&A...516A..84Z}
{Zaqarashvili}, T.~V., {D{\'{\i}}az}, A.~J., {Oliver}, R., \& {Ballester},
  J.~L. 2010, \aap, 516, A84

\bibitem[{{Zaqarashvili} {et~al.}(2011){Zaqarashvili}, {Khodachenko}, \&
  {Rucker}}]{2011A&A...529A..82Z}
{Zaqarashvili}, T.~V., {Khodachenko}, M.~L., \& {Rucker}, H.~O. 2011, \aap,
  529, A82

\bibitem[{{Zhugzhda} \& {Goossens}(2001)}]{2001A&A...377..330Z}
{Zhugzhda}, Y.~D. \& {Goossens}, M. 2001, \aap, 377, 330

\end{thebibliography}
\end{document}